\documentclass[11pt,a4paper]{article}

\usepackage{graphicx}
\usepackage{amssymb,amsmath}
\usepackage{xcolor}
\usepackage{booktabs}
\usepackage{tabularx}
\usepackage{placeins}
\usepackage{authblk}

\begin{document}


\title{Data-driven RANS closures for wind turbine wakes under neutral conditions}
\author{Julia Steiner \and Richard P. Dwight \and Axelle Vir\'{e}}
\date{\today}
\maketitle




\begin{abstract}
The state-of-the-art in wind-farm flow-physics modeling is Large Eddy Simulation (LES) which makes accurate predictions of most relevant physics, but requires extensive computational resources.  The next-fidelity model types are Reynolds-Averaged Navier-Stokes (RANS) which are two orders of magnitude cheaper, but resolve only mean quantities and model the effect of turbulence.  They often fail to accurately predict key effects, such as the wake recovery rate.  Custom RANS closures designed for wind-farm wakes exist, but so far do not generalize well: there is substantial room for improvement.  In this article we present the first steps towards a systematic data-driven approach to deriving new RANS models in the wind-energy setting.  Time-averaged LES data is used as ground-truth, and we first derive optimal corrective fields for the turbulence anisotropy tensor and turbulence kinetic energy (t.k.e.) production.  These fields, when injected into the RANS equations (with a baseline $k-\varepsilon$ model) reproduce the LES mean-quantities.  Next we build a custom RANS closure from these corrective fields, using a deterministic symbolic regression method to infer algebraic correction as a function of the (resolved) mean-flow.  The result is a new RANS closure, customized to the training data.  The potential of the approach is demonstrated under neutral atmospheric conditions for multi-turbine constellations at wind-tunnel scale.  The results show significantly improved predictions compared to the baseline closure, for both mean velocity and the t.k.e.\ fields.
\end{abstract}


\maketitle

\section{Introduction}\label{sec:intro}

Offshore wind farms have the potential to become the sustainable future power plants of North-Western Europe. For instance, the Dutch government projects a growth towards $11.5\,\mathrm{GW}$ of installed offshore capacity in 2030, entailing that a large part of the North Sea will be filled with wind farms \cite{rvo2020}.

Accurate wind turbine wake models are important because they facilitate optimizing energy yield and turbine loading during the design and operation phases of a wind farm. There exist a multitude of models that attempt to model wake effects, varying in physical fidelity, accuracy, and computational cost.  They range from simple engineering models to complex computational fluid dynamics codes \cite{stevens2017}. Generally, engineering models are not accurate enough if significant wake interaction is present. The state of the art is Large Eddy Simulations (LES) which is a high-fidelity Computational Fluid Dynamics (CFD) method where most of the scales of turbulence are resolved whilst the smallest scales are modeled. However, this type of simulation requires extensive computational resources: one wind speed and direction simulation of the Lillgrund wind farm can take between $160k$ and $3000k$ processor hours depending on how the turbines are modeled \cite{breton2017,ghaisas2017}. The next-fidelity model types are Reynolds-Averaged Navier-Stokes (RANS) models which require about two orders of magnitude less computational time (because they model all turbulence scales) resulting in coarser meshes and direct equations for the mean quantities - without a need for time-averaging the simulation. Of course, sometimes also transient quantities are of interest and the atmospheric boundary layer is inherently transient. Nonetheless, RANS models provide useful information for time-averaged quantities over short intervals.  For both RANS and LES the range of scales present, ranging from the boundary-layer on the turbine blades to the height of the atmospheric boundary layer, is too large to be fully resolved.  Generally, actuator models are used to model the presence of wind turbines \cite{sanderse2011}.

In this article, we aim to extend the capabilities of RANS turbulence models for wind turbines under quasi-steady conditions. Currently, the most commonly used RANS model in this setting, namely the $k-\varepsilon$ model, has crippling structural shortcomings.  It over-predicts the eddy viscosity in the near wake which leads to an over-prediction of the wake recovery, and it fails to account for the effects of turbulence anisotropy \cite{sanderse2011}. There are two main reasons why the model does not perform well in the near wake: (i) the eddy-viscosity assumption is invalid in the near wake region, and (ii) the direct effect of the turbine on the turbulence mean quantities is not modeled \cite{rethore2009}.  Several modifications to the baseline $k-\varepsilon$ model have been proposed in the literature. Most approaches aim at extending the baseline Linear Eddy Viscosity Model (LEVM) by either adding additional terms to the transport equations or by directly adding an eddy-viscosity limiter in the momentum equation.  For example, El Kasmi and Masson \cite{elkasmi2008} used a modified version of the $k-\varepsilon$ model which introduces an additional source term that is proportional to the square of the turbulent kinetic energy production rate in the transport equation for $\varepsilon$. The source term is only non-zero close to the rotor because they argue that this is the area where non-equilibrium effects are important. This source term is intended to suppress the overproduction of turbulent kinetic energy in the near wake where strong shear gradients are present. Prospathopoulos et al. \cite{prospathopoulos2011} apply an eddy-viscosity limiter (Durbin limiter) based on a realizability constraint.  R\'{e}thor\'{e} \cite{rethore2009} used two different eddy-viscosity limiters based on a realizability constraint, and the adverse pressure gradient in the near wake region. Van der Laan et al. \cite{laan2015} developed a model named the $k-\varepsilon-f_P$ model with a limiter that reduces the eddy-viscosity in regions with high-velocity gradients. The limiter is a simplified version of a cubic non-linear eddy viscosity model (NLEVM) and is applied directly in the relation for the eddy viscosity. In a follow-up publication, van der Laan et al. \cite{laan2018} compare this eddy-viscosity limiter to the one from Shih and Durbin, all for the $k-\varepsilon$ model. They recommend the use of either the $f_P$ or the Shih limiter since the Durbin limiter is very sensitive to ambient turbulence levels.

Full NLEVMs have also been used in turbine wake modeling: Gomez-Elvira et al. \cite{gomez2005} and van der Laan et al. \cite{laan2013} used a NLEVM which yielded improved predictions of velocity and Reynolds stresses.  However, the models they devised showed numerical instabilities for high turbulence intensities and fine meshes. Cabezon et al. \cite{cabezon2011} went further, using a Reynolds Stress Model (RSM) which again improved predictions at the cost of robustness.

While all of these models offer some improvements over the standard $k-\varepsilon$ model, the improvements are test-case dependent, some of them require turbine and/or test-case specific tuning parameters, some of them are not numerically robust and at this point, the influence of atmospheric stratification is not considered.  Further, most of these models aim only at improving the shortcomings of the eddy-viscosity assumption, they do not directly consider the effect of actuator forcing on the turbulence equations.

In this publication we take a different approach to improve the unsatisfactory baseline model, namely this work aims to further develop the data-driven framework introduced by Schmelzer et al.\  \cite{schmelzer2019} Sparse Regression of Turbulent Stress Anisotropy (SpaRTA). This framework has, until now, only been applied to simple 2d testcases with Reynolds numbers below $50,000$.  Data-driven turbulence modelling is a recent development in the fluid-dynamics community and its merit has generally been resticted to relatively simple two-dimensional flows \cite{ling2016,durbin2018,duraisamy2019,xiao2019,Kumar_2020}.  Data-driven approaches to turbulence modeling can be divided into two broad categories based on the underlying regression model: either using (a) extremely general models with a very large number of parameters, such as artificial neural networks and random forests \cite{Kaandorp_2020,tracey2015,parish2016,singh2016,singh2017,ling2016}; or (b) and methods using symbolic algorithms such as sparse regression and Gene Expression Programming (GEP) which tend to result in concise, inspectable models~\cite{schmelzer2019,weatheritt2016,weatheritt2017,weatheritt2020,zhang2020customized}.
While the former ``black-box'' approaches were the first to be applied to turbulence modelling, and are capable of capturing very complex models, the result is expensive to incorporate into a CFD solver, and often makes the solver highly unstable.  As a result most such models are used as a corrective step rather than as true turbulence closures~\cite{ling2016,Kaandorp_2020}.  Symbolic algorithms avoid these pitfalls due to the simplicity and comprehensibility of the resulting expressions.

A critical aspect of data-driven modelling is the location in the governing equations at which the baseline model is modified: some authors scale the turbulent kinetic energy production term in the LEVM \cite{tracey2015,duraisamy2019}; others introduce a correction to the anisotropy tensor, thereby transforming an LEVM into a NLEVM \cite{ling2016,Kaandorp_2020}.  In this publication, we use both an anisotropy correction and a source term in the transport equation for the turbulent kinetic energy, following SpaRTA \cite{schmelzer2019}. This has the benefit of correcting both the directionality and the magnitude of the Reynolds stress tensor (RST), as well as accounting for model-form errors in the transport equation for $k$.  Furthermore SpaRTA uses deterministic symbolic regression, for which the search space is constrained towards parsimonious algebraic models using modern sparsity-promoting regression techniques \cite{brunton2016,rudy2017}.

In terms of novelty, the authors know of only two examples of data-driven turbulence modeling applied to wind farms, namely those from Adcock et al.~\cite{adcock2018} and King et al.~\cite{king2018}. The papers employ quite a different approach to us, and do not go beyond a two-dimensional model.  Adcock et al.\ use an adjoint approach in a 2D RANS solver with a mixing length turbulence model.  They fit optimal mixing length and thrust coefficient fields, which are generalized into a closure using a Gaussian mixture model.  King et al.\ use a similar adjoint-based approach where they directly solve for an optimal eddy viscosity field and then use Gaussian progress regression to parametrize the correction beyond the training dataset.

In this work, we use the SpaRTA framework and employ it to find an improved $k-\varepsilon$ model using time-averaged LES data from several two- and three-turbine constellations at the wind-tunnel scale under neutral ABL conditions.  From an application point-of-view, this represents a relatively limited data-set -- from which it is nevertheless possible to derive a novel RANS closure that significantly improves the predictive capability of the baseline model, generalizing will between the constellations.  This is the first work to apply sparse regression for turbulence modelling to high-Reynolds number, three-dimensional problems -- demonstrating the potential of SpaRTA and similar methods for industrial problems.

The publication is structured as follows: in Section \ref{sec:methodology} we define the entire methodology of the approach. Additive model-form error terms within the $k-\varepsilon$ LEVM model are identified via the introduction of corrections to the stress-strain relation and the turbulence transport equations. The $k$-corrective-frozen-RANS approach to identify the optimal corrections is explained.  Then, the generalization of these correction terms using an elastic net is introduced. In Section \ref{sec:results}, the results of the frozen approach, the training, and cross-validation of the correction terms and the inclusion of the correction terms in the flow solver are displayed. Finally, conclusions are drawn in Section \ref{sec:conclusions}.

\section{Methodology}\label{sec:methodology}
Our complete data-driven turbulence modeling chain, consists of three main steps.  First we define test-cases and perform LES simulations, to provide training and validation data.  This data serves as a target and ground-truth of the subsequent RANS modelling efforts (Section~\ref{s:les}).  Secondly we solve for RANS corrective fields (Section~\ref{s:frozen}).  There are fields, which when injected into a RANS simulation of the training-cases, reproduce the LES mean-field and turbulence intensity.  Note that it is not sufficient to merely use the LES-obtained Reynolds-stress tensor (RST) to correct the momentum equation, as established by Thompson et al.~\cite{thompson2016} this does not necessarily lead to the correct mean-flow.  Rather, our procedure serves the same purpose as {\it field inversion} in the work of Parish \& Duraisamy~\cite{parish2016}, but does not require an adjoint or an optimization.  Thirdly, we use sparse symbolic regression to discover a concise algebraic expression approximating these corrective fields, using only local flow quantities available in the RANS simulation (Section~\ref{s:sparsereg}).  The result is a new turbulence closure model, customized to the training test-cases, which can be used to make predictions for similar cases outside the training set.

\subsection{Test-case definition and LES database generation}
\label{s:les}
\begin{figure}[h]
\centering
\includegraphics[clip,trim={0.2\textwidth} {0.1\textwidth} {0.2\textwidth} {0.1\textwidth},width=0.6\textwidth]{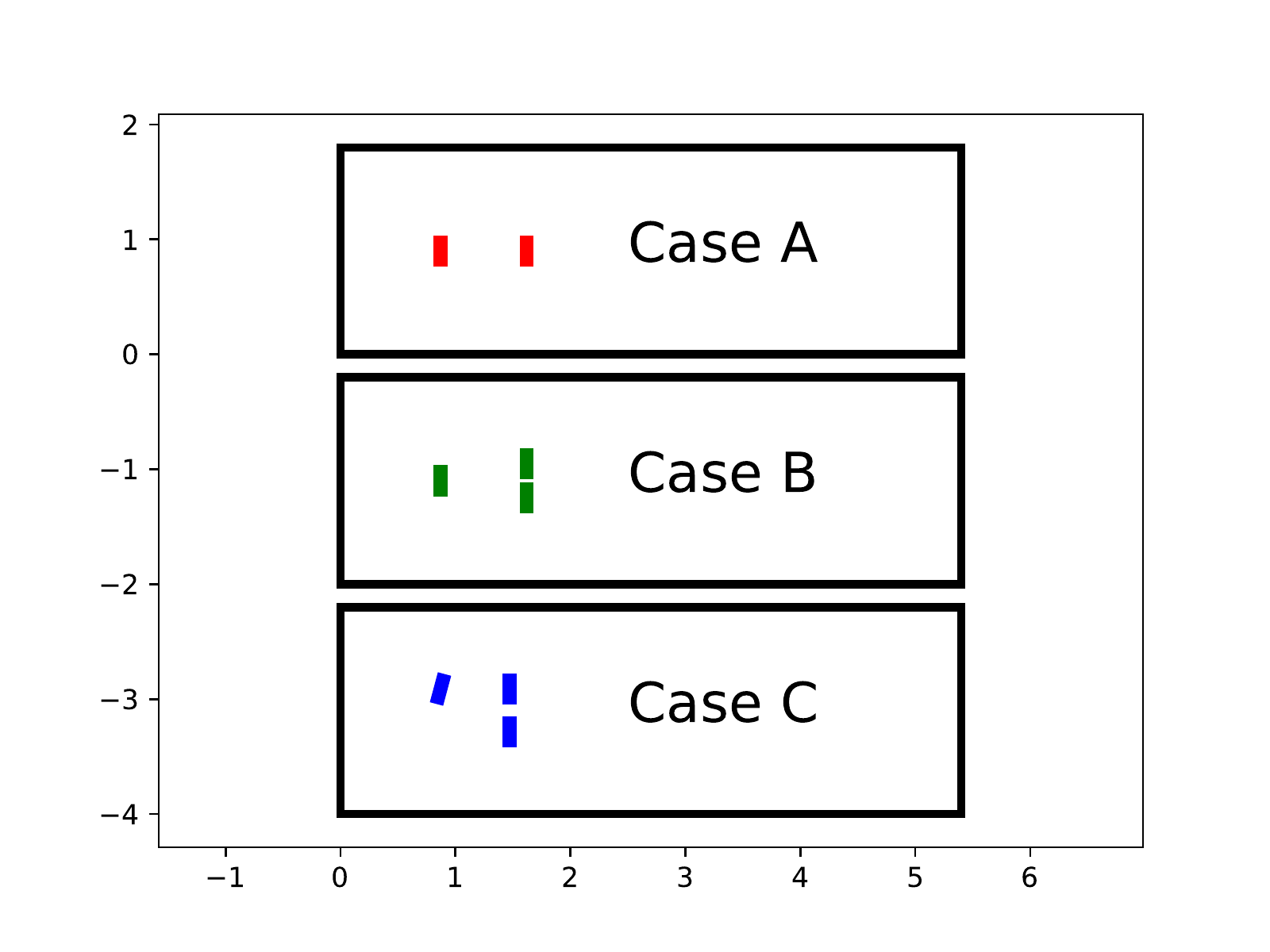}
\caption{Case constellation, turbine diameter is to scale.}\label{fig:cases}
\end{figure}

The first step in the proposed methodology is to set up a database of cases that serve as a ground-truth, to both train- and validate new closure models.  For this publication, the database consisted of three different cases. The same surface roughness and hub-height velocity were used for all cases, but the turbine constellation was changed, as visualized in Figure~\ref{fig:cases}. The turbine and inflow properties correspond to the wind-tunnel experiment from Chamorro and Port\'{e}-Agel \cite{chamorro2010}, the most important parameters are listed in Table~\ref{tab:cases}.
%


\begin{table}[h]
\centering
\begin{minipage}{18pc}
\centering
\begin{tabular}{@{}*{7}{l}}
\toprule \midrule
\textbf{Turbine} & \\
\midrule
Diameter&$D=0.15\text{m}$\\
Hub height&$h_{hub}=0.125\text{m}$\\
Rotation speed &$\Omega=1190\text{rpm}$\\
\midrule
\textbf{Inflow boundary} & \textbf{layer} \\
\midrule
Velocity &$U\left(h_{hub}\right)=2.2\text{m/s}$\\
Turbulence intensity &$\sigma_U\left(h_{hub}\right) =1.0 \%$\\
\midrule
\textbf{Mesh} & \\
\midrule
Domain size &  $5.4 \times 1.8 \times 0.46 \text{m}^3$ \\
Resolution & $360 \times 120 \times 64 $ \\
\midrule \bottomrule
\end{tabular}\caption{Case setup paramters}\label{tab:cases}
\end{minipage}
\begin{minipage}{18pc}
\centering
\begin{tabular}{@{}*{7}{l}}
\toprule \midrule
\textbf{WALE model} & \\
\midrule
$C_e$ & $0.93$\\
$C_k$ & $0.0673$\\
$C_w$ & $0.325$\\
\midrule
\textbf{$k-\varepsilon$ model} &  \\
\midrule
$C_\mu$ & $0.03$\\
$C_{\varepsilon 1}$ & $1.42$\\
$C_{\varepsilon 2}$ & $1.92$\\
$\sigma_{\varepsilon}$ & $1.3$\\
$\sigma_{k}$ & $1.3$\\
\midrule \bottomrule
\end{tabular}\caption{Turbulence model parameters}\label{tab:turbParameters}
\end{minipage}
\end{table}

For the CFD model, OpenFOAM-6.0 is used in conjunction with the SOWFA-6 toolbox \cite{sowfa}. For the RANS solver, a modified $k-\varepsilon$ model is the baseline closure; for the LES solver, the WALE model is used to model the unresolved scales \cite{nicoud1999,sanz2017}. The closure coefficients used here for the two models can be found in Table~\ref{tab:turbParameters}. Validation of both turbulence models is carried out on the benchmark case from Chamorro and Porte-Ag\'{e}l. Additionally, Xie and Archer's results \cite{xie2015} are used to determine an appropriate mesh resolution for the LES simulations. SOWFA's actuator disc model with the same turbine geometry, rotational speed and force projection parameter is used in both the RANS and LES simulations. The turbine geometry is detailed in Stevens et al. \cite{stevens2018}. No controller is used in the simulations, the turbine is run at a fixed rotational speed $\Omega$.  For the force projection, the Gaussian width is chosen to be twice the largest cell size in the rotor area $\epsilon=0.03\text{m}$ \cite{martinez2012}. For simplicity and to avoid interpolation errors, the same mesh resolution was used for both RANS and LES throghout the majority of the paper, though in practice the RANS simulations would typically be run at a lower resolution. At the end of the publication in Section \ref{sec:meshConvergence}, a mesh convergence study is carried out by varying the mesh density of the baseline and the corrected RANS simulations. The ABL is modelled in the LES by means of a precursor simulation with doubly periodic boundary conditions, and a uniform body-force applied to achieve the desired hub height velocity.  A zero-flux condition was used at the top of the domain for both the precursor and the simulations with turbines.  In the latter, periodic boundary conditions were used at the sides, a zero-gradient boundary condition at the outlet, and at the inlet plane instantaneous fields from the precursor are applied.  At the ground, standard boundary conditions for a rough wall are used, see Section~\ref{sec:bcWall}. For both RANS and LES, second-order discretization schemes are used in space with the exception of the convection terms in the turbulence transport equations for the RANS model where an first-order upwind scheme is used for numerical stability. The temporal discretization for the LES simulations is a second order Crank-Nicolson scheme.

\begin{figure}[h]
\includegraphics[width=0.24\textwidth]{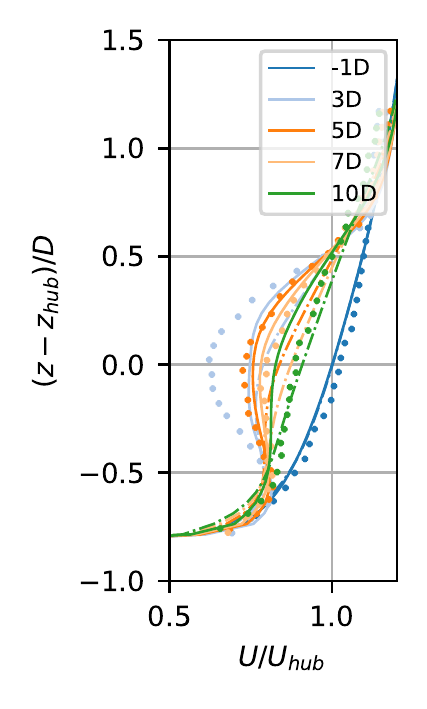}
\includegraphics[width=0.24\textwidth]{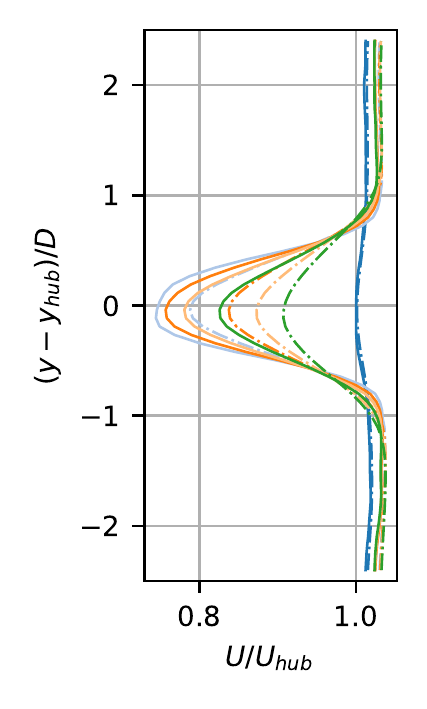}
\includegraphics[width=0.24\textwidth]{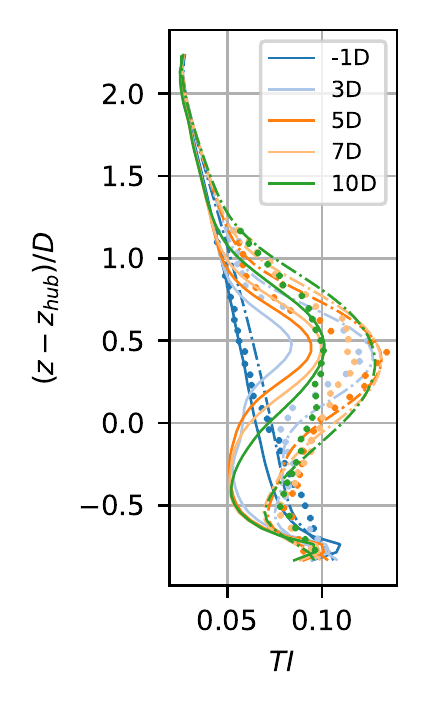}
\includegraphics[width=0.24\textwidth]{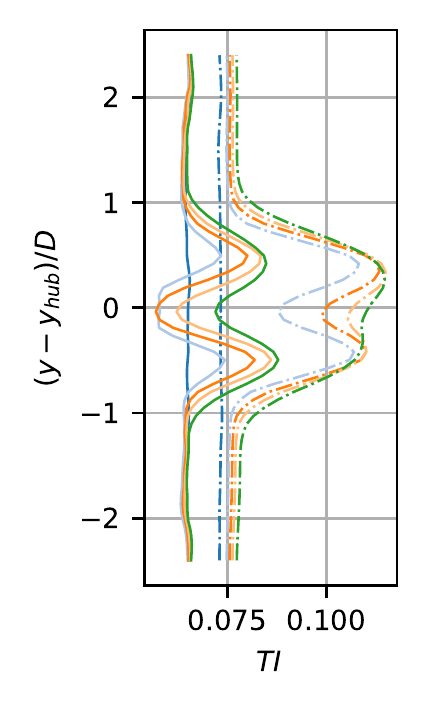}
\caption{\label{fig:validation} Validation of case setup and turbulence models through vertical and horizontal slices of the flow field up and downstream of the rotor plane in terms of velocity and turbulence intensity, solid line corresponds to LES, dash dotted line corresponds to RANS, and dots belong to experimental results.}
\end{figure}

Figure \ref{fig:validation} shows the validation of the models on the benchmark case in terms of mean velocity and turbulence intensity. As expected RANS over-predicts turbulence intensity and wake recovery as compared to LES. Nevertheless, neither one of the models perfectly matches the experiment, possibly also due to the relatively low Reynolds number of the wind tunnel setup (the wall functions and the RANS turbulence model are derived for higher Reynolds numbers). The boundary layer height $\delta$ based Reynolds number of the wind tunnel experiment is $\text{Re}_\delta = U_\infty \delta /\nu \approx 930,000$ \cite{chamorro2010}. Further, the LES simulations show an unphysical overshoot in the turbulent kinetic energy close to the wall. The peak in the turbulent kinetic energy in the LES simulations is a well documented problem for LES simulations with wall functions for rough walls \cite{brasseur2010}. This is something that can be improved in future publications. The authors would like to stress that the aim of this publication is not to perfectly reproduce the experiments, but to showcase the potential of a methodology that systematically improves RANS based predictions using time-averaged LES data.

\FloatBarrier

\subsection{Discovery of optimal corrective fields}
\label{s:frozen}
Given LES reference data for a given test-case, we aim to find corrections to the RANS equations in the form of frozen fields for that same test-case, such that RANS matches time-averaged LES in terms of mean velocity and turbulence intensity.  The core approach presented here was developed by Schmelzer et al. \cite{schmelzer2019}, to which we add two modifications specific to the wind-farm application.

The method is similar to the ``frozen approach'' for estimating turbulence dissipation rates from LES results.  Namely variables that are known from LES are injected into model equations, and the values of the remaining variables are deduced.  Specifically, let LES quantities be denoted by a $\star$, so the LES mean velocity is $U^\star$, turbulent kinetic energy $k^\star$ and Reynolds stresses $\tau_{ij}^\star$, whereby both resolved and SGS modeled turbulence quantities are implied.  Let the baseline $k-\varepsilon$ model be modified in two places: (i) in the momentum equation add a correction to the normalized anisotropy tensor, denoted  $\textcolor{red}{\tilde{b}_{ij}^\Delta}$, and (ii) in the equation for $k$ add a scalar correction term $\textcolor{blue}{\tilde{R}}$ which accounts for errors in the turbulent kinetic energy production and other inconsistencies in the transport equation for the turbulent kinetic energy. These correction terms are both spatially varying fields (tensor and scalar respectively), and are embedded in the model as:
\begin{equation}\label{eq:kEpsilonABLk}
        \frac{D k^\star}{D t} = \mathcal{P}_k^\star + \textcolor{blue}{\tilde{R}}
         - \varepsilon +\frac{\partial }{\partial x_j} \left[  \left(\nu +\nu_t/\sigma_k \right) \frac{\partial {k^\star}}{\partial x_j} \right],
\end{equation}
\begin{equation}\label{eq:kEpsilonABLepsilon}
        \frac{D \varepsilon}{D t} = \left[C_{\varepsilon 1} \left(\mathcal{P}_k^\star+
                                        \textcolor{blue}{\tilde{R}} \right)
                                         -C_{\varepsilon 2} \varepsilon  \right] \cdot                                             \frac{\varepsilon}{k^\star}
                                         +\frac{\partial }{\partial x_j} \left[  \left(\nu + \nu_t/\sigma_\varepsilon\right) \frac{\partial \varepsilon}{\partial x_j} \right]
\end{equation}
where the production term is defined as
\begin{equation}
\mathcal{P}_k^\star := 2 k^\star b_{ij}^\star \frac{\partial U_i^\star}{\partial x_j}
\end{equation}
with
\begin{equation}
\label{eq:bij}
b_{ij}^\star := \frac{\tau_{ij}^\star}{2 k^\star} - \frac{1}{3} \delta_{ij} = -\frac{\nu_t}{k^\star} S_{ij}^\star + \textcolor{red}{\tilde{b}_{ij}^\Delta}.
\end{equation}

Given an initial guess for $\varepsilon$ (e.g. from the baseline $k-\varepsilon$ model), $\textcolor{blue}{\tilde{R}}$ can be computed directly from \eqref{eq:kEpsilonABLk}.  Subsequently $\varepsilon$ can be updated by solving \eqref{eq:kEpsilonABLepsilon} with the most recent $\textcolor{blue}{\tilde{R}}$, and we iterate back and forth until convergence.  Then $\textcolor{red}{b_{i j}^\Delta}$ can be computed directly from \eqref{eq:bij} and the definition of the eddy-viscosity:
\begin{equation}
\nu_t := C_\mu \frac{{k^\star}^2}{\varepsilon}.
\end{equation}
The resulting fields satisfy the modified $k-\varepsilon$ equations, with the LES data as a solution.

In practice, two adjustments are made to this procedure to address issues specific to the wind-farm application: (i) blending of the correction terms to zero at the bottom and the top of the domain, and (ii) an atmospheric boundary-layer correction which only varies in the direction perpendicular to the wall.

\subsubsection*{Blending of the turbulence correction terms} The blending term at the top and the bottom of the domain is introduced to avoid interaction between the correction terms and the boundary conditions. The blending term $F_\beta$ employed in this publication is a simplified version of the one used by Menter \cite{menter1994} for the blending of the $k-\varepsilon$ and the $k-\omega$ model into the $k-\omega$ \textit{SST} model. It is formulated as
\begin{equation}
F_\beta(z) =
\begin{cases}
\tanh \left[\left(
\frac{z}{z_{\mathrm{lower},\beta}}
\right)^ \alpha \right] \ \ \ \ \ \ \ \ \ \ \textbf{for $z \leq z_{\mathrm{mid}}$}\\
\tanh \left[\left(
\frac{z_\mathrm{max}-z}{z_\mathrm{max}-z_{\mathrm{upper},\beta}}
\right)^ \alpha \right] \ \ \  \textbf{for $z>z_{\mathrm{mid}}$}\\
\end{cases}
\end{equation}
where the exponent $\alpha$ determines how fast the blending transitions between $0$ and $1$, $\beta \in \{\mathrm{ABL},\mathrm{wake}\}$ is used to distinguish between the different blending applied to the correction terms for the ABL and for the main simulation, $z_{\mathrm{mid}}$ and $z_{\mathrm{max}}$ are related to the domain dimensions, and finally, $z_{\mathrm{lower},\beta}$ and $z_{\mathrm{upper},\beta}$ are domain specific threshold parameters. In \cite{menter1994} the lower bound for the blending is chosen according to the nondimensional wall distance. But since in this publication a relatively simple case with uniform surface roughness and flat terrain is used, this is not necessary.  Generally, different blending terms can be used for all the correction terms. However, in this particular case, using two different blending functions between the ABL and the wake correction worked well. The  parameters used here are found in Table~\ref{tab:blending}. The wall blending for the ABL corrections was chosen such that the correction is zero in the first cell center.
\begin{table}
  \centering
   \begin{tabular}{cc}
   \toprule \midrule
   Parameter & Value \\
   \midrule
$\alpha$ & $4$\\
$z_\mathrm{mid}$ & $0.23\text{m}$\\
$z_\mathrm{max}$ & $0.46\text{m}$\\
$z_\mathrm{upper,all}$ & $0.4\text{m}$\\
   \midrule \bottomrule
   \end{tabular}
   \qquad
  \begin{tabular}{cc}
  \toprule \midrule
  Parameter & Value \\
  \midrule
$z_{\mathrm{lower},ABL}$ & $0.01\text{m}$\\
$z_{\mathrm{lower},\mathrm{wake}}$ & $0.05\text{m}$\\
  \midrule \bottomrule
  \end{tabular}
  \caption{Blending parameters for the blending function $F_\beta$.}
  \label{tab:blending}
\end{table}

\subsubsection*{Matching RANS boundary-layer profiles to LES}\label{sec:bcWall}
In the undisturbed ABL, LES and the baseline RANS model give different profiles for mean-velocity and turbulent kinetic energy.  Even though the LES precursor profile is set as the RANS inflow, it evolves before contact with the turbines.  As such, if the profiles are not matched, the RANS corrective fields that are discovered will necessarily include some component that corrects the ABL mismatch, and some other component to correct the turbine wake.  We prefer to separate these corrections, and so first match the ABL profiles. To achieve this two modifications are applied: (i) the boundary condition representing the ground for the two simulations is made consistent, and (ii) the velocity profiles away from the boundaries are adjusted through a one-dimensional RANS closure correction varying as a function of wall normal distance only.

Matching the boundary condition at the wall is complicated by the use of wall models in both LES and RANS.  In particular both use equilibrium assumptions and the log-law for a rough wall to determine skin-friction.  They assume that first cell is in the log-layer, so that e.g.
\[
\tau_{xz} \simeq -\rho u_\star^2
\]
but they estimate $u_\star$ differently.  In the LES, the time averaged velocity at the first cell $U_1$ at height $z_1$ above the wall, is used to estimate an average friction velocity, using
\[
u^\mathrm{LES}_\star \simeq \frac{\kappa U_1}{\log(z_1/z_0)},
\]
where $\kappa$ is the Karman constant and $z_0$ is the roughness height.  The local instantaneous wall friction is then computed using a Schumann boundary condition.  On the other hand, RANS relates the local friction velocity to the turbulent kinetic energy in the first cell
\[
u^\mathrm{RANS}_\star \simeq \sqrt[4]{C_\mu}\sqrt{k_1}
\]
and then uses the log-law to determine an expression for the eddy viscosity there.  For consistency we require that in the RANS boundary-condition
\[
C_\mu = \left(u^\mathrm{LES}_\star\right)^4/k_1^2,
\]
and we choose $C_\mu = 0.055$ to satisfy this equation based on the LES precursor.

The model parameter $C_\mu$ appears also in the definition of the eddy viscosity and it has a large influence on the turbulent kinetic energy.  In fact, in this role it can be used to regulate the turbulence intensity at hub height.  In the standard $k-\varepsilon$ model \cite{launder1974} the recommended value is $C_\mu=0.09$, but for atmospheric boundary layers a value of $C_\mu=0.03$ is often suggested \cite{sogachev2012}. As a consequence, in the remainder of this work, the baseline RANS simulations will use $C_\mu=0.03$, and the corrected simulations $C_\mu=0.055$.

In addition in RANS, we used the standard equilibrium-assumption boundary-condition for epsilon:
\[
\varepsilon = \frac{C_\mu^{3/4} k^{3/2}}{\kappa z_0}.
\]

Having matched boundary conditions between RANS and LES, the profiles of $U$ and $k$ still do not match sufficiently well.  The frozen approach described above is applied using the LES precursor as a data source, and a RANS simuation of a flat-plate.  Two corrections for the ABL $\textcolor{red}{b_{ij}^{\Delta,ABL}}$ and $\textcolor{blue}{R^{ABL}}$ are obtained that eliminate remaining differences almost everywhere.

\begin{figure}[h]
\centering
\includegraphics[clip,trim={0.05\textwidth} {0.} {0.12\textwidth} {0.},width=\textwidth]{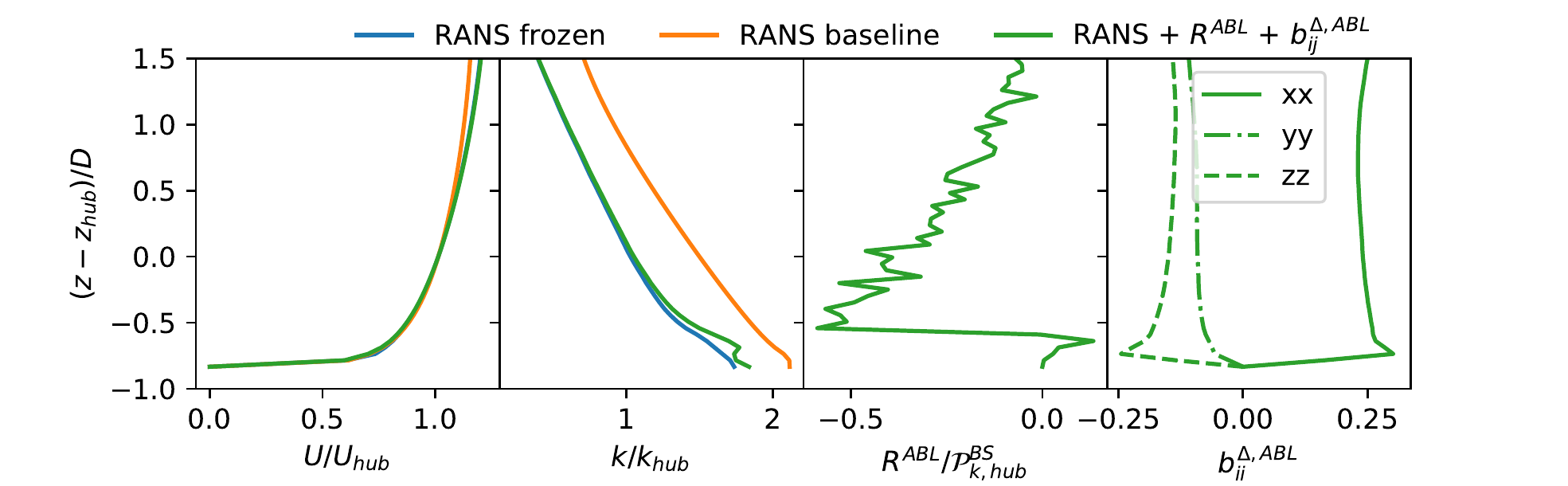}
\caption{\label{fig:fsCorrection} Matching of ABL profiles (denoted free-stream ``FS'') between the frozen RANS, the baseline RANS and, the corrected RANS simulations. The one-dimensional profiles for the velocity, the turbulent kinetic energy and the two correction terms are shown. The scalar correction term is normalized with the Boussinesq turbulent kinetic energy production in the \textcolor{red}{free-stream.}}
\end{figure}

Figure~\ref{fig:fsCorrection} shows the resulting profiles from the frozen approach and then the profiles in case the corrections are propagated. For the propagated cases the domain forcing is chosen such that the hub height velocity matches. The velocity profiles between the frozen and the propagated RANS simulation match very well, but the turbulent kinetic energy profiles do not match well close to the wall. In fact, the unphysical overshoot in the turbulent kinetic energy is also observed in the propagated RANS simulations, even though the peak was removed from the LES reference data. However, the turbulent kinetic area in the rotor wake matches well between LES and corrected RANS, and this is what is relevant for this publication.

\subsubsection*{Full formulation of correction terms}
Finally, now the full formulation for the correction terms can be written as
\begin{align}
\textcolor{blue}{\tilde{R}} &=
  F_{\mathrm{wake}}    \cdot \textcolor{blue}{R}
+ F_{ABL} \cdot \textcolor{blue}{R^{ABL}} \\
\textcolor{red}{\tilde{b}_{ij}^\Delta} &=
  F_{\mathrm{wake}}    \cdot \textcolor{red}{b_{ij}^\Delta}
+ F_{ABL} \cdot \textcolor{red}{b_{ij}^{\Delta,ABL}}
\end{align}
with blending terms $F_\beta$, ABL correction terms $R^{ABL}$, $b_{ij}^{\Delta,ABL}$ and wake correction terms $R$, $b_{ij}^{\Delta}$. In the next section, generalized expressions for the wake correction terms are inferred. Contrary to the wake correction terms, the ABL correction terms can be used as-is. However, this means that they are not general and need to be recomputed if one of the following parameters changes: surface roughness, inflow velocity, and - depending on how strong Coriolis effects are - wind direction.

\subsection{Learning of correction terms}
\label{s:sparsereg}
Once the optimal wake correction terms $\textcolor{blue}{R}$ and $\textcolor{red}{b_{ij}^\Delta}$ are known, a generalized expression for the correction terms is inferred using a deterministic symbolic regression method similar to the one presented in Schmelzer et al. \cite{schmelzer2019}.  For this, the generalized nonlinear eddy viscosity formulation as proposed by Pope \cite{pope1975} is used.  Assuming that the anisotropy tensor depends only on the local strain rate $S_{ij} = \frac{1}{2} \left( \partial_j U_i+\partial_i U_j\right)$ and rotation rate tensors $\Omega_{ij} = \frac{1}{2} \left( \partial_j U_i-\partial_i U_j\right)$, then an almost perfectly general mapping has the form
\begin{equation}
\textcolor{red}{b^\Delta_{ij}} \left( S_{ij}, \Omega_{ij} \right) =
\sum_{n=1}^{10} T_{ij}^{(n)} \alpha_n \left( I_1,\dots,I_5 \right)
\label{eq:bdelta}
\end{equation}
where $T_{ij}^{(n)}$ are integrity basis tensors, and $\alpha_n:\mathbb{R}^5\rightarrow\mathbb{R}$ are ten, arbetrary scalar-valued functions of the five invariants $I_n$.  In practice, invariants of $\textbf{S}$ and $\boldsymbol\Omega$ prove insufficient to represent the required correction $b^\Delta$~\cite{Kaandorp_2020}.  Therefore in the following we use invariants of the set $\{\textbf{S},{\boldsymbol \Omega},\textbf{A}_p,\textbf{A}_k\}$ where $\textbf{A}_p = -I \times \nabla p$ and $\textbf{A}_k = -I \times \nabla k$, thereby making pressure- and $k$-gradients available to the model, see e.g.~\cite{wang2017}.  The resulting, already large, feature-space is suplemented with additional nondimensional scalar features listed in Appendix~\ref{app:physicalFeatures}, including -- for instance -- actuator forcing.  So, finally each $\alpha_n$ in this work is potentially a function of $F\simeq 60$ features.

An analogous modelling approach is taken for the correction term $\textcolor{blue}{R}$, now explicitly using the $F$ features $I_1,\dots,I_F$:
\begin{equation}
\textcolor{blue}{R} \left( S_{ij}, \Omega_{ij} \right) = 2 k \frac{\partial u_i}{\partial x_j}
\sum_{n=1}^{10} T_{ij}^{(n)} \alpha_n^R \left( I_1,\dots,I_F \right)
+ \varepsilon \cdot \beta^R \left( I_1,\dots,I_F \right).
\end{equation}
The main difference being that two types of terms are used: one that mirrors a correction to the turbulence production and one that represents a more general scalar correction. The scalar term is scaled with the turbulent dissipation $\varepsilon$ to keep the term dimensionally correct, and note that the immediate vicinity of the wall is excluded from the training by $F_\mathrm{wake}$.  The motivation for these two seperate corrections is to capture both errors in the production term itself (which in many flows is the dominant term), as well as other model-form errors, notably the omission of the effect of the rotor forcing on the turbulence.  A detailed derivation of this missing terms can be found in Rethore \cite{rethore2009} and will not be repeated here.   They dependend on the local values for the rotor forcing, the velocity fluctuations, and the pressure gradient, and do not have the shape of a production term.

The task of the supervised machine-learning algorithm, is then to find a suitable formulation for the functions $\alpha_n$, $\alpha_n^R$, and $\beta^R$ based on the dataset available from LES, including the corrections discovered by the frozen approach in the previous section.  This is now a standard regression problem, albeit with a high-dimensional input space and large number of data-points (determined by the total number of mesh-points in the training LES simulations).  We consider many well-known machine-learning methods unsuitable for this problem, notably both artificial neural networks and random-forests, produce complex nonsmooth models, are prone to overfitting, expensive to evaluate, and unlikely to be well-trained in the entire relevant subspace (given limited training data).  Rather we prefer regression techniques that ensure relatively simple closed-form expressions (see for example the models discovered in the next section, \eqref{eq:R} and \eqref{eq:bDelta}).

In this work we use sparse regression \cite{schmelzer2019}.  The $F$ input features are used to build a large library of $L$ candidate (basis) functions $(\ell_1, \dots \ell_L)$. This is done by recombining features with each other (up to a maximum of three features), and applying exponentiation by $-1$,$-\frac{1}{2}$, $\frac{1}{2}$, and $2$.  This already results in a library exponentially larger then $F$.  The scalar functions are then represented as:
\begin{equation}
  \alpha_n(\textbf{I}) \simeq \sum_{k=0}^L \theta^n_k \ell_k(\textbf{I}).
  \label{eq:alpha}
\end{equation}

In principle, a least-squares approach (possibly $L^2$-regularized) could be used here to find the coefficients $\hat\Theta$.  However, this yields massively complex models prone to overfitting the data.  Practically, due to multi-collinearity in the input data, large coefficients $\theta$ result with cancellation between terms.  Such models are unsuitable for implementation in CFD models as they are numerically stiff.  Hence, a sparsity promoting approach is used here in which sets most of the $\theta_k$ to zero, this is achieved by a loss-function with a both $L^1$- and $L^2$-regularization, i.e.\ an elastic net~\cite{Zou_2005}.  However the number of features and data-vector is large enough that feature- and library reduction is required as a preliminary step.  Note that both \eqref{eq:bdelta} and \eqref{eq:alpha} are linear in the coefficients $\hat\Theta \in \mathbb{R}^{10\cdot L}$, and therefore for a given data-set of size $N$ we can find a matrix $C \in \mathbb{T}^N\times \mathbb{R}^{10\cdot L}$ (where $\mathbb{T}$ is here the space of $3\times 3$ tensors), such that \eqref{eq:bdelta} and \eqref{eq:alpha} are encapuslated by
\[
\mathbf{b}^\Delta = C \Theta.
\]

The outline of the procedure is as follows:
\begin{enumerate}
\item{\textbf{Preprocessing}} using a mutual information to eliminate features, building the library, and then {\it cliqueing} (identifying sets of colinear functions) to reduce the library.
\item{\textbf{Model discovery}} using an elastic net to identify important library functions.  By varying regularization parameters $\lambda$ and $\rho$, the result is an array of models with a variety of complexity and accuracy.  The optimization problem is:
\begin{equation}
\min_{\Theta} \left[
\left\| C {\Theta} - \textbf{b}^\Delta \right\|_2^2 +
\lambda \rho \left\| {\Theta} \right\|_1 +
0.5 \lambda (1-\rho) \left\| {\Theta} \right\|_2^2 \right]
\end{equation}
where $\textbf{b}^\Delta$ is the target anisotropy correction at the $N$ mesh-points.
\item \textbf{Remove unnecessary functions} from the library by eliminating all basis functions for which the corresponding $\theta=0$ for each of the models found in (ii).  The matrix $C \rightarrow \tilde{C}$ and $\Theta \rightarrow \tilde{\Theta}$ are also reduced.
\item{\textbf{Model calibration}} using Ridge regression to identify the magnitude of the model coefficients for the previously derived array of corrections.  Again a regularization parameter $\lambda_R$ is used to encourage small coefficients:
\begin{equation}
\min_{{\Theta}} \left[
\left\| \tilde{C} {\Theta} - \textbf{b}^\Delta \right\|_2^2 +
\lambda_r \left\| {\Theta} \right\|_2^2 \right]
\end{equation}
\end{enumerate}

I.e. in steps (iii) and (iv), the sparsity information from the elastic net is retained, and the coefficient values are discarded.  A more detailed description of the approach is given by Schmelzer et al.~\cite{schmelzer2019}.

The preprocessing step is a novelity with respect to \cite{schmelzer2019}, necessitated by the increased feature set and data size used here.  Mutual information (MI) between the input features and the correction terms was calculated {\it a priori} to determine if a feature was relevant for the regression~\cite{Goderie2020}.  This involved treating the data-set as samples from random variables describing the features and the corrections, and estimating their MI using a kernel density estimator, following~\cite{Moon_1995,miSource}.  MI has the advantage of measuring the amount of information (in bits) obtained about the output, given an observation of the input.  As such it does not rely on any model (linear or otherwise), or any assumptions about the form of the random variables.  This is in contrast to e.g.\ correlation coefficients, which by their linearity assumption are wholely unsuited for feature selection for nonlinear models.

The {\it cliqueing} procedure is motivated by the high multi-colinearity obtained by default within the library.  Specifically we compute the correlation coefficient between all pairs of library functions, and then group them by finding {\it cliques} whose correlation with each other all exceeds the cut-off of $0.99$.  Finding cliques is an established problem in graph theory~\cite{cliqueSource}.  We then select the algebraically simplest member of the clique to represent the clique, and discard the remainder, knowing that the data is in any case insufficient to distiguish them.   Here -- in contrast to MI -- a linear measure of correlation is adequate, because members of the candidate library are combined linearly in \eqref{eq:alpha}.

Finally, since the models are simple, explicit expressions for the correction terms, they can be directly integrated into the RANS solver.

\section{Results and discussion}\label{sec:results}
This section shows the application of the proposed methodology to the previously described dataset. Resulting flow fields with the optimal and the learned correction terms are shown and discrepancies are discussed in detail.

\subsection{Flow field with optimal correction terms}
\begin{figure}[h]
\centering
\includegraphics[clip,trim={0.05\textwidth} {0.} {0.12\textwidth} {0.},width=\textwidth]{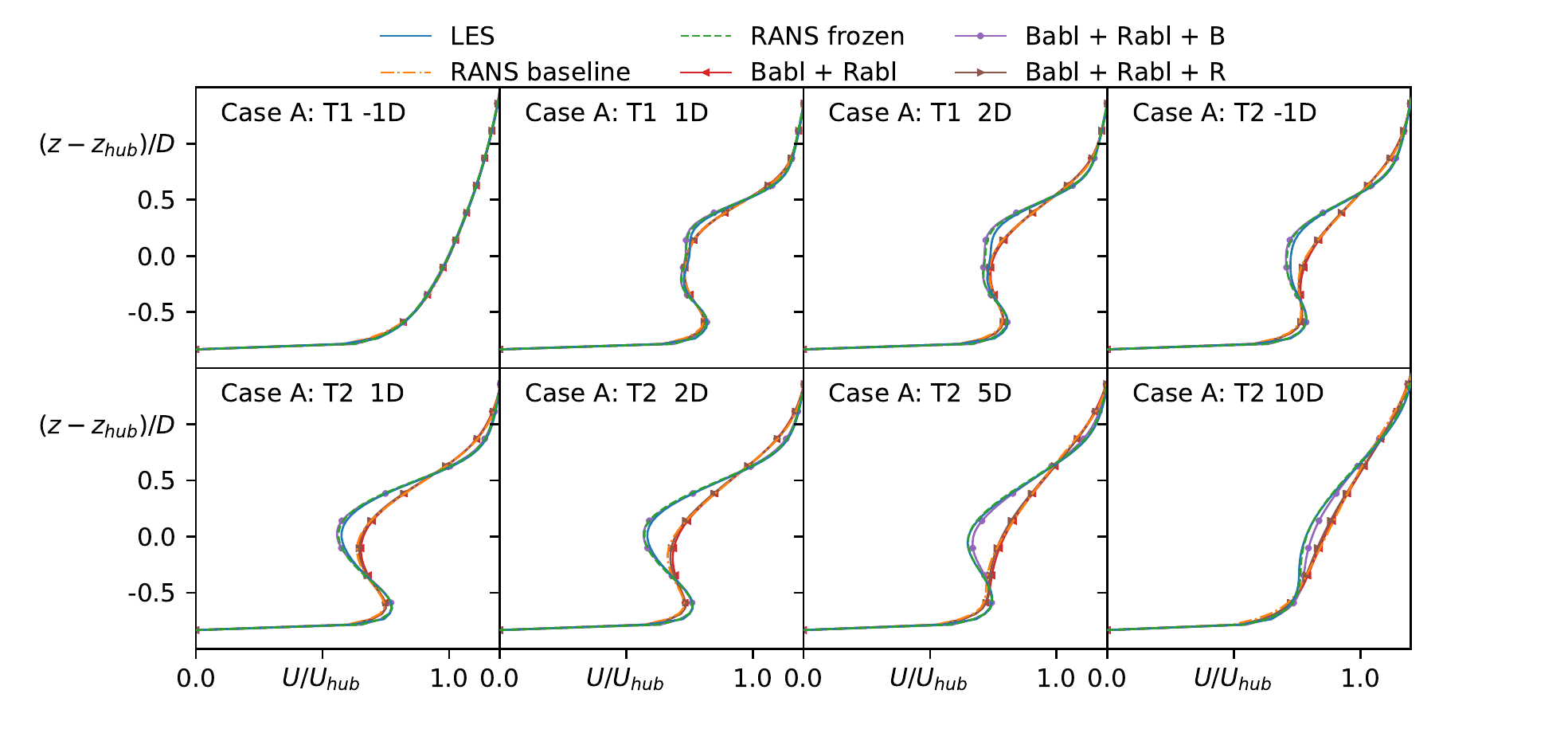}
\caption{\label{fig:modelFrozenU} Comparison between LES, RANS baseline, and frozen RANS with selective inclusion of the different components of the (frozen) correction terms.  Vertical slices of the velocity field up and downstream of the rotor plane of the two turbines of case A.}
\end{figure}

\begin{figure}
\includegraphics[clip,trim={0.05\textwidth} {0.} {0.12\textwidth} {0.},width=\textwidth]{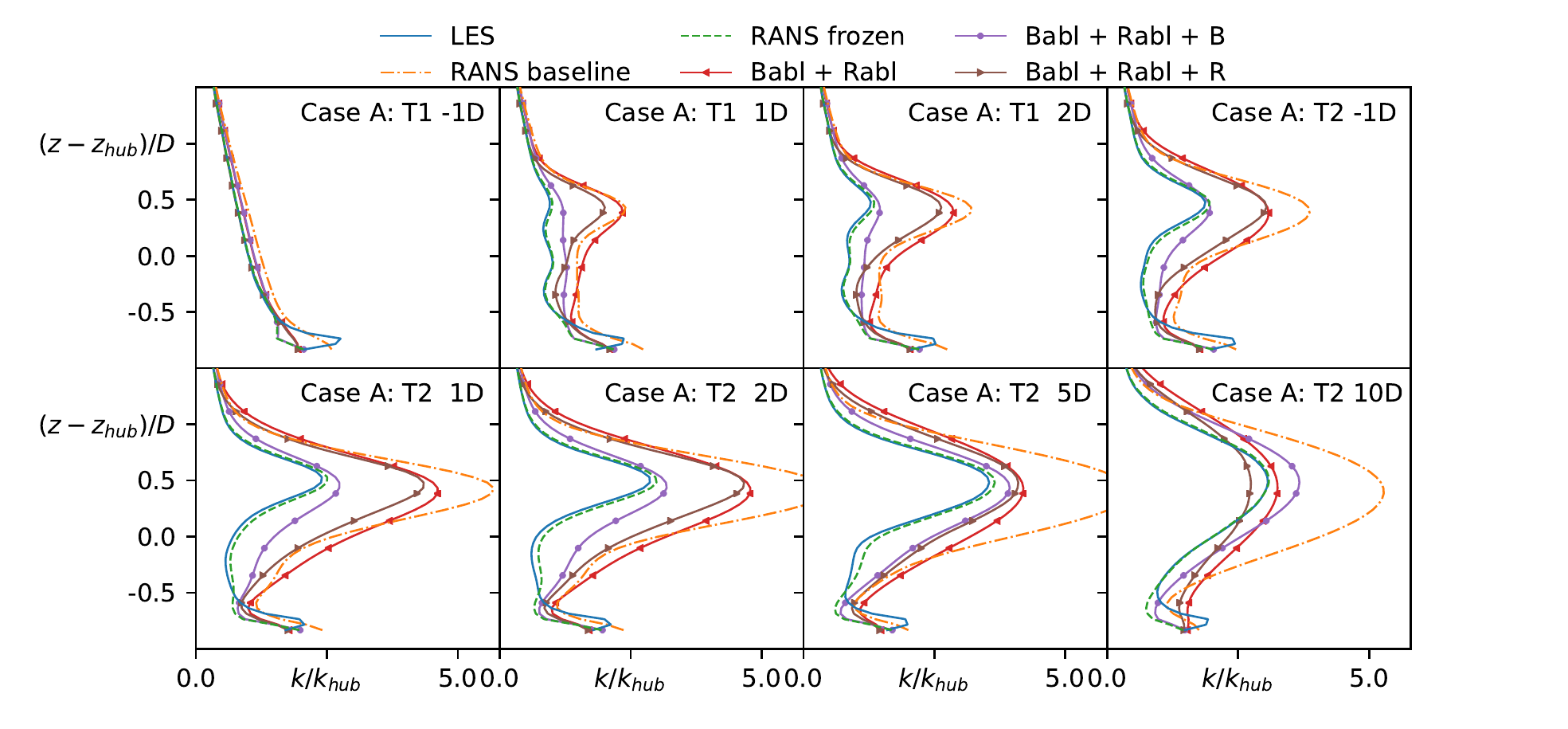}
\caption{\label{fig:modelFrozenK} Comparison between LES, RANS baseline, and frozen RANS with selective inclusion of the different components of the (frozen) correction terms.  Vertical slices of the turbulent kinetic energy field up and downstream of the rotor plane of the two turbines of case A.}
\end{figure}

The optimal correction terms are derived for the three cases in the dataset. Subsequently, the (static) optimal corrections are integrated into the RANS turbulence models for all the three test cases. The results obtained from this are referred to as ``frozen'' or ``optimally corrected'' RANS. Figures \ref{fig:modelFrozenU} and \ref{fig:modelFrozenK} show the wake development as predicted by the LES, the baseline RANS, and the frozen RANS simulations using vertical slices through the flow field. The horizontal slices can be found in the appendix in Figures \ref{fig:modelFrozenUhor} and \ref{fig:modelFrozenKhor}. Optimally corrected RANS represent the best-case scenario that can be obtained when using this methodology. In the next subsection, the generalized models for the correction terms will introduce additional errors. The results in the figure show that indeed the optimal correction terms lead to an almost perfect match between LES mean and frozen RANS velocity and turbulent kinetic energy fields.

The relative importance of the different frozen correction terms for the prediction of the velocity and turbulent kinetic energy field is also visible from Figures \ref{fig:modelFrozenU} and \ref{fig:modelFrozenK}, respectively. Some conclusions can be drawn from the selective inclusion of the correction terms. The free-stream corrections $R^{ABL}$ and $b_{ij}^{\Delta,ABL}$ do not have much effect on the velocity field, but they slightly reduce the overprediction of the turbulent kinetic energy. Of course, this is closely tied to the choice of $C_\mu$. The anisotropy correction term $b_{ij}^\Delta$ is more important than the scalar correction terms $R$. If only a correct prediction of the velocity field is necessary, then the scalar term $R$ can be neglected. However, the scalar correction term $R$ does yield some improvement for the prediction of the turbulent kinetic energy over the case where only the tensor correction term $b_{ij}^\Delta$ is used.

\FloatBarrier

\subsection{Learning of correction terms}

The results presented in the following are based on the datasets for constellation A \& C as presented in Figure \ref{fig:cases}. Case A is used for the training of the models and case C is used to cross-validate the learned correction terms. Note that case C is more complex than case A because it includes one more turbine and one of the turbines is yawed with respect to the incoming flow. The training dataset does not include the entire dataset of case A, rather only entries centered around the turbines wake are used. This helps avoid overfitting and reduced the dataset somewhat. The exact criteria for inclusion in the training dataset are $x_{rotor}-1.0 D < x < x_{rotor}+20.0 D$, $y_{rotor}-1.5 D < y < y_{rotor} + 1.5 D$, and $0.05\text{m} < z < z_{rotor}+1.5 D$.

\newcolumntype{b}{X}
\newcolumntype{s}{>{\hsize=.8\hsize\centering\arraybackslash}X}
\newcolumntype{x}{>{\hsize=.2\hsize\centering\arraybackslash}X}

\begin{table}[!htbp]
    \centering
	\begin{tabularx}{0.95\linewidth}{x b s s}
    \toprule \midrule
        ID & Description & Raw feature & Normalization\\ \midrule
  $q_{\gamma}$ & Shear parameter & $\left\|\frac{\partial U_i}{\partial x_j}\right\|$
  		& $\frac{\varepsilon}{k}$ \\
  $q_{\tau}$ & Ratio of total to normal Reynolds stresses &  $||\overline{u_i'u_j'}_{Boussinesq}||$    &  $k$ \\
  $q_{\nu}$   &   Viscosity ratio &   $\nu_t$    &   $100\nu$ \\
  $q_{TI}^{{\dagger}}$   &   Turbulence intensity  &   $k$    &   $\frac{1}{2}U_iU_i$ \\
  $q_{F}^{{\dagger}}$ & Actuator forcing
  		& $\left\|F_{cell}\right\|$ & $\frac{1}{2} \rho_0 A_{cell} \left\| U \right\|^2$ \\
	\midrule \bottomrule
	\end{tabularx}
    \caption{List of non-dimensionalized physical features used in the model discovery phase and their precise definition. The features that are not Galilean invariant are marked with ${\dagger}$.}
    \label{tab:physicalFeatures}
\end{table}

Following the methodology outlined previously, the feature set used to construct a library of basis functions is based on the results of the mutual information analysis between features and correction terms. A list of the input feature set divided into physical parameters and invariants that were obtained as a result of the preprocessing step can be found in Tables \ref{tab:physicalFeatures} and \ref{tab:invariants}, respectively. The full list of physical features and invariants used as an input to the mutual information algorithm can be found in the Appendices \ref{app:physicalFeatures} and \ref{app:integrityBasis}. Additionally, only the first four tensors of the integrity basis are used where $T^{(1)} = S$, $T^{(2)} = S \Omega - \Omega S$, $T^{(3)} = \text{dev} \left(S^2 \right)$,
$T^{(4)} = \text{dev} \left(\Omega^2 \right)$ and $\text{dev}$ is the deviatoric part of the tensor.  Applying the cliqueing algorithm to the library of basis functions that was constructed from the reduced feature set, further reduced the size of the library by around a factor of 6.

\begin{table}
  \centering
   \begin{tabular}{cc}
   \toprule \midrule
   Invariant ID & Definition \\
   \midrule
   $I_1$ & $\textbf{S}^2$  \\
   $I_2$ & $\boldsymbol{\Omega}^2$  \\
   $I_{19}$ & $\boldsymbol{\Omega} \textbf{A}_k \textbf{S}^2$  \\
   $I_{25}$ & $\textbf{A}_k^2 \textbf{S} \boldsymbol{\Omega} \textbf{S}^2$  \\
   $I_{35}$ & $\textbf{A}_p \textbf{A}_k \textbf{S}^2$  \\
   \midrule \bottomrule
   \end{tabular}
   \qquad
  \begin{tabular}{ccc}
  \toprule \midrule
  Tensor ID & Definition & Normalization \\
  \midrule
  $\textbf{S}$
    & $\frac{1}{2} \left( \frac{\partial u_i}{\partial x_j}+\frac{\partial u_j}{\partial x_i}\right)$
    & $\frac{\varepsilon}{k}$\\
  $\boldsymbol{\Omega}$
    &  $\frac{1}{2} \left( \frac{\partial u_i}{\partial x_j}-\frac{\partial u_j}{\partial x_i}\right)$
    & $\frac{\varepsilon}{k}$\\
  $\textbf{A}_k$
    & $-I \times \nabla p$
    & $\frac{\varepsilon}{\sqrt{k}}$ \\
  $\textbf{A}_p$
    & $-I \times \nabla k$
    & $\rho_0 \left\| u \nabla u \right\|$\\
  \midrule \bottomrule
  \end{tabular}
  \caption{List of invariants used in the model discovery phase and their precise definition.}
  \label{tab:invariants}
\end{table}

\begin{figure}[h]
\centering
\includegraphics[clip]{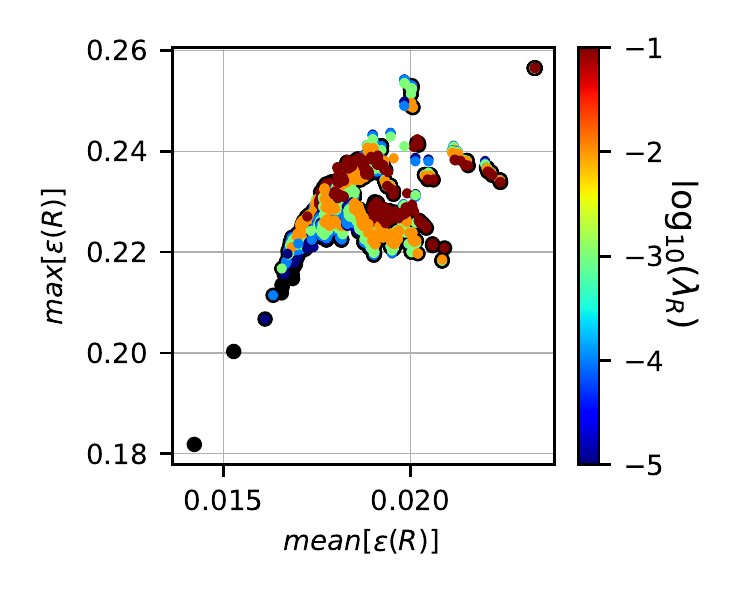}
\includegraphics[clip]{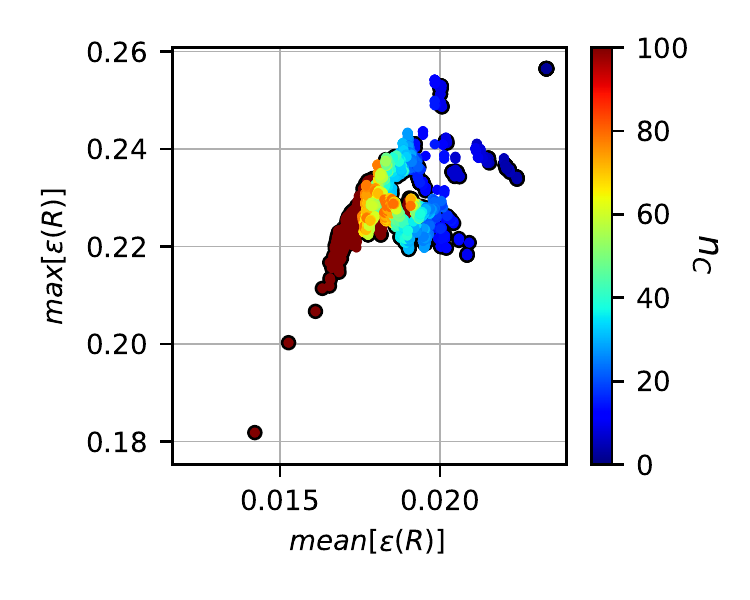} \\
\includegraphics[clip]{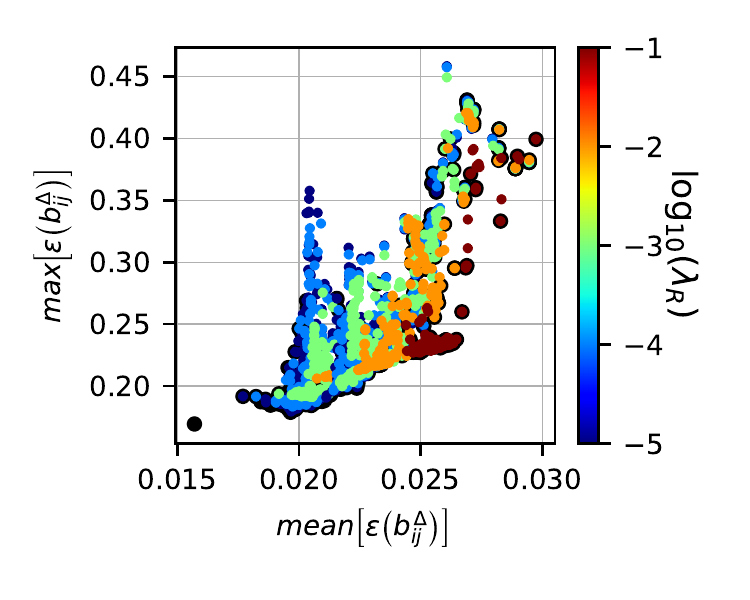}
\includegraphics[clip]{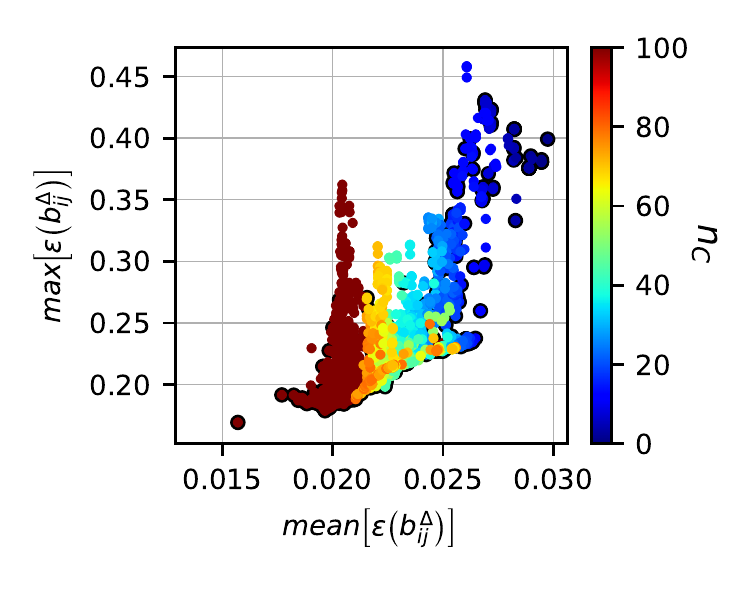}
\caption{\label{fig:modelDisovery} Scatter plot of all the models obtained for both correction terms. Members of the three-dimensional Pareto front with respect to mean and max error, as well as model complexity are highlighted in black. The coloring of the elements is according to the magnitude of the Ridge $\lambda_R$ penalization parameter and the model complexity $n_C$ . }
\end{figure}

Subsequently, the three-step regularization methodology is applied to determine, first which candidate functions are important, and secondly what the magnitude of these coefficients should be.  The result is a very large number of potential models: Figure \ref{fig:modelDisovery} shows the results of this process for both the anisotropy correction $\textcolor{red}{b_{ij}^\Delta}$ and the scalar correction term $\textcolor{blue}{R}$. The left side of the figure illustrates the trade-off between the anticipated robustness and the model accuracy by showing the influence of the Ridge regularization parameter $\lambda_R$ on the mean and maximum error of the model on the training data set. The right side of the figure visualizes the trade-off between the model complexity and the model accuracy by highlighting the number of terms of the model. But the results are not straight-forward and only limited trends can be identified. In general, more complex models are seen to give better predictions for both correction terms, but this is not always the case. The trend with respect to an increasing regularization parameter $\lambda_R$ is different for the two correction terms. For the anisotropy correction $b_{ij}^\Delta$ higher regularization correlates with a higher mean error but a lower maximum error. For the scalar correction term $R$ higher regularization generally leads to both higher mean and maximum error.

Because the model discovery and calibration phase generate a lot of models and because it was was difficult to pick which models should to be selected for further investigation, the three-dimensional Pareto front in terms of mean error, maximum error, and model complexity was computed.  This is indicated in Figure~\ref{fig:modelDisovery} by black outlines. Going forward only the models which are a member of the Pareto front are investigated. Since the number of Pareto optimal models is still of the order of around $500$, a further automated selection procedure is necessary.  Cliqueing was again applied, this time to predictions of complete models, and models that were too similar were discarded.

\begin{figure}[h]
\centering
\includegraphics[clip,trim={0.0\textwidth} {0.} {0.1\textwidth} {0.},width=\textwidth]{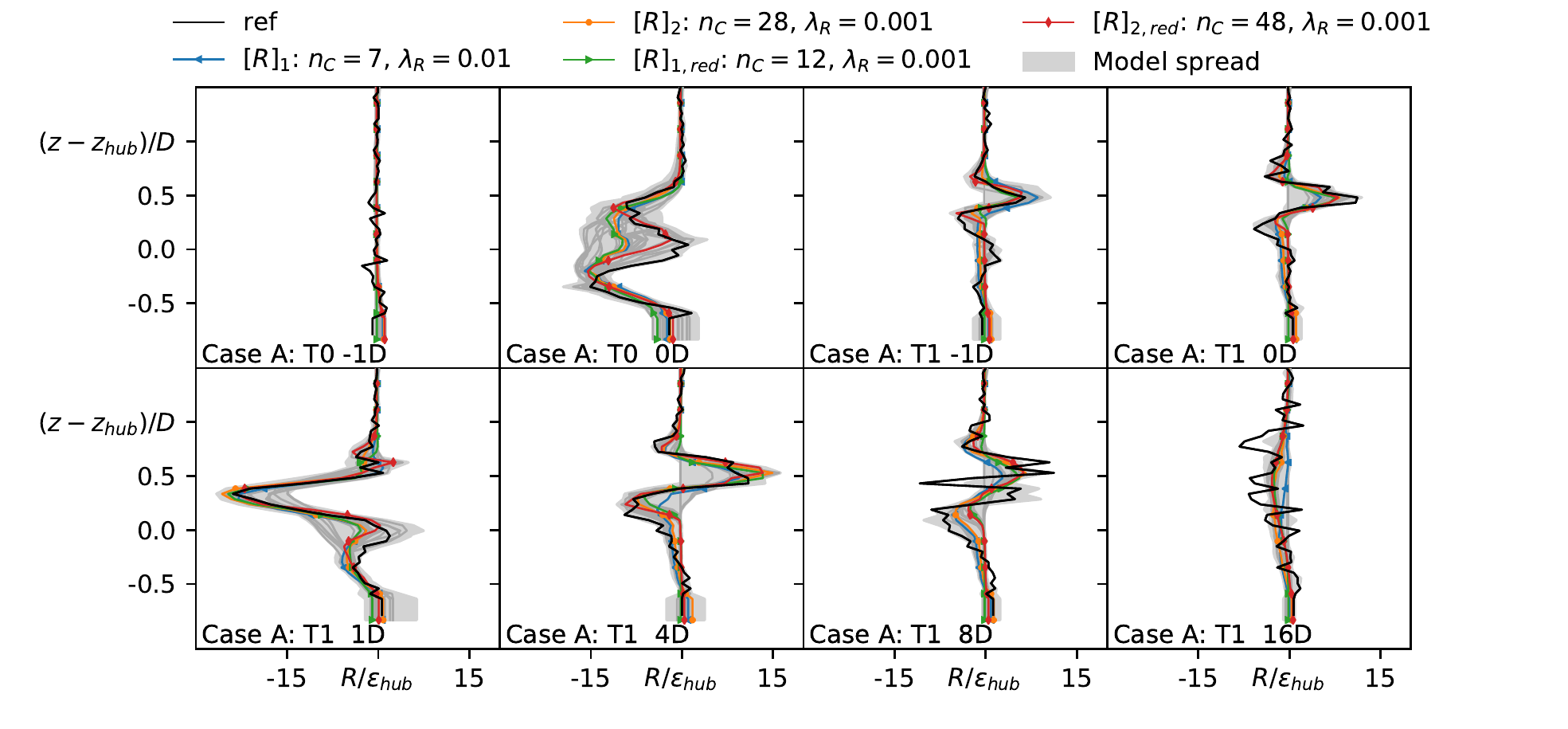}
\caption{\label{fig:modelSpreadR} Spread of trained models for $R$ for Case A.  Vertical slices at the rotor plane at different streamwise stations.  The model spread is for all models that are Pareto optimal. The models selected during the cliqueing post-processing step are shown explicitly either in color or in dark gray. The models selected for further investigation are highlighted in color. Finally, the optimal correction term is shown in black.}
\end{figure}

\begin{figure}[h]
\centering
\includegraphics[clip,trim={0.0\textwidth} {0.} {0.05\textwidth} {0.},width=\textwidth]{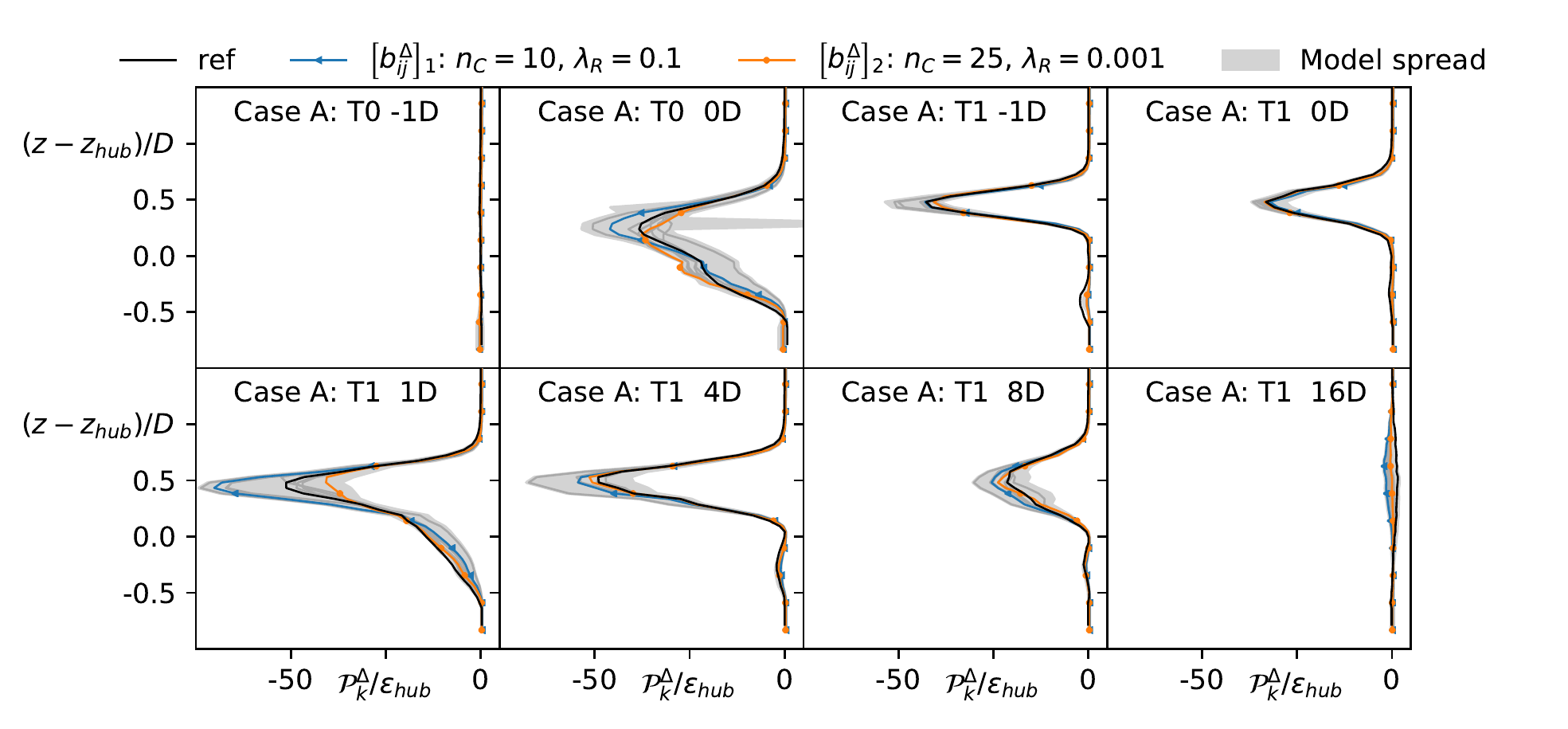}
\caption{\label{fig:modelSpreadPkDelta} Spread of trained models for $b^\Delta$, visualized using $\mathcal{P}_{k}^\Delta$ for case A.  Vertical slices at the rotor plane at different streamwise stations.  The model spread is for Pareto optimal models.  The models selected during the cliqueing post-processing step are shown explicitly either in color or in dark gray. The models selected for further investigation are highlighted in color. Finally, the optimal correction term is shown in black. }
\end{figure}

The effect of this procedure is visualized in Figures \ref{fig:modelSpreadR} and \ref{fig:modelSpreadPkDelta} for $R$ and $b^\Delta$ respectively.  The figures show the spread of Pareto-optimal models, the subselection of models obtained from the cliqueing, and finally the models selected for implementation in the CFD solver. The anisotropy correction is visualized by means of its effect on the turbulent kinetic energy production  $\mathcal{P}_k^\Delta = 2 k b_{ij}^\Delta \frac{\partial u_i}{\partial x_j}$.  Our experience shows that this is a good indicator for the accuracy of the anisotropy correction term, and substantially easier to visualize.

Figure \ref{fig:modelSpreadR} shows the predictions for the scalar correction term $R$. It can be seen that the entire selection of the spread of models can be reduced to about 20 models. The four models highlighted in color are the ones that will be implemented in the CFD solver in the next section. The four models were selected based on accuracy and complexity. Further, also the models named $\left[R\right]_1$ and $\left[R\right]_2$ contain terms with both negative and positive powers of the input features, whereas the models named $\left[R\right]_{1,red}$ and $\left[R\right]_{2,red}$ only contain terms with positive powers. This was done because -- unsurprisingly -- the negative powers negatively affected the convergence of the models once implemented in the CFD model. For the $R$ term this effect was not always present, but for the anisotropy correction term $b_{ij}^\Delta$ none of the models including negative powers lead to convergence.  As such = they are not discussed further.

Figure \ref{fig:modelSpreadPkDelta} shows the spread of the model prediction for the anisotropy prediction. Again, with a reduced set of about 10 models, the entire spread of results can be covered. Two models were selected for further investigation as a trade-off between accuracy and complexity.  All of the models contain only positive powers of the input features. Going forward the two selected models will be referred to as $\left[b_{ij}^\Delta\right]_1$ and $\left[b_{ij}^\Delta\right]_2$.

\subsection{Robustness of correction terms}
The correction-learning methodology employed in this publication is completely decoupled from the CFD model. This significantly simplifies the regression as compared to an online approach where the terms are trained while coupled with the CFD model.  However, this also means that once a coupling with the CFD solver is constructed, the correction terms may not be the same as predicted during the learning stage. Further, at this point, there is no clear criteria or methodology to determine the stability of a correction model {\it a priori}.  Hence simple testing and cross-validation is the most immediate strategy.

Experience with the framework has shown that models that are very complex, i.e. above about 50 terms, tend not to converge for either one of the correction terms.  Furthermore, in models trained on our data, the Ridge regression parameter should be at least $\lambda_R \geq 0.001$ to assure convergence not only on the training but also on test datasets.
%
\begin{figure}[h]
\centering
\includegraphics[clip,trim={0.0\textwidth} {0.} {0.0\textwidth} {0.},width=\textwidth]{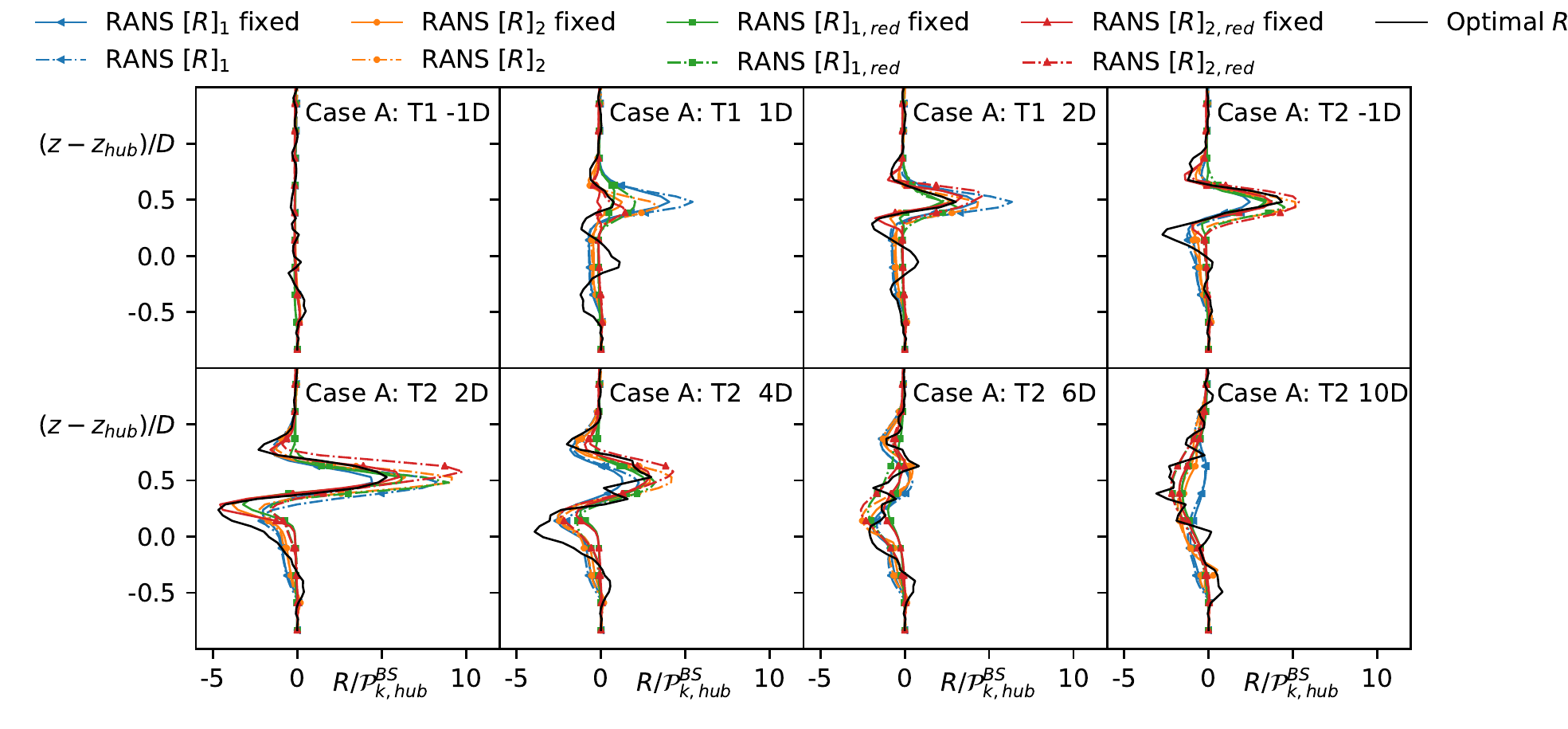}
\caption{\label{fig:robustnessR} Spread of fixed and coupled corrections $R$ for the training case A via vertical slices at the rotor plane up and downstream of the two turbines.}
\end{figure}

\begin{figure}[h]
\centering
\includegraphics[clip,trim={0.0\textwidth} {0.} {0.04\textwidth} {0.},width=\textwidth]{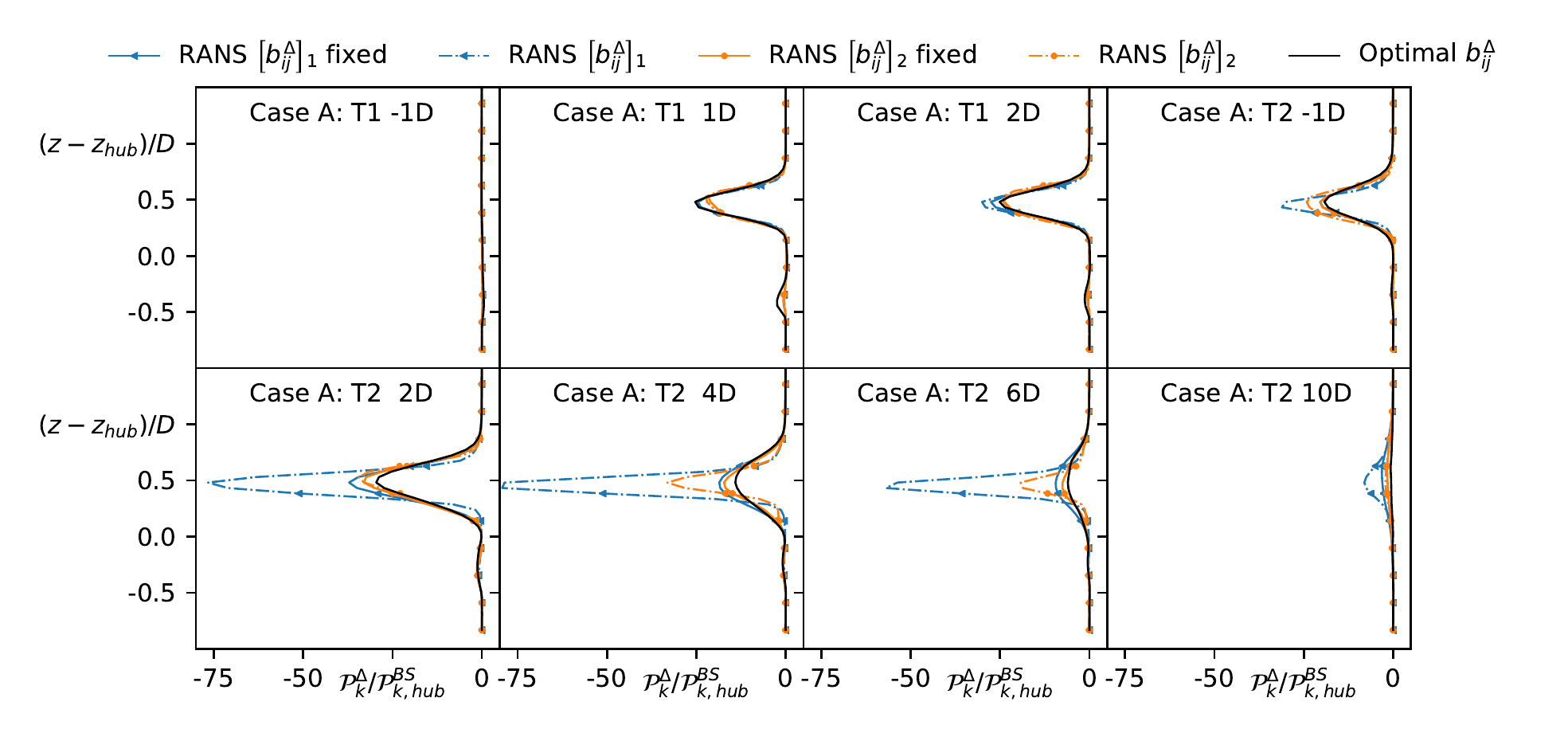}
\caption{\label{fig:robustnessPkDelta} Spread of fixed and coupled corrections $b_{ij}^\Delta$ in terms of  $\mathcal{P}_{k}^\Delta$ for the training case A via vertical slices at the rotor plane up and downstream of the two turbines.}
\end{figure}

To make the assessment of the models structured, the robustness of the two correction terms is assessed separately before implementing both terms simultaneously in the turbulence model.  For example, the robustness of the model for $R$ can be assessed by using the frozen correction for $b_{ij}^\Delta$, and vice versa.  In Figures \ref{fig:robustnessR} and \ref{fig:robustnessPkDelta} the robustness of the previously selected correction terms is shown for the scalar and the anisotropy correction models.

Figure \ref{fig:robustnessR} compares the optimal correction terms for $R$ with the one obtained when coupled with the CFD model and the fixed one obtained during the learning phase with no coupling to the CFD solver. Ideally, the coupled and the fixed term would overlap perfectly. However, as visible from the figure, this is not the case and the effect is more or less pronounced for the different pictured correction models. The discrepancy between the coupled and the fixed terms is larger in regions where the optimal term has high gradients. If the discrepancy between the two terms is too large, the model not only becomes inaccurate but may also lead to an unstable coupling once both correction terms are introduced simultaneously.
%

Figure \ref{fig:robustnessPkDelta} shows the same analysis for the two selected model for the anisotropy correction $b_{ij}^\Delta$ in terms of the modified turbulent kinetic energy production term $\mathcal{P}_k^\Delta = 2 k b_{ij}^\Delta \frac{\partial u_i}{\partial x_j}$. Again, in regions where the optimal correction term and its derivative are large, the disparity between the fixed and the coupled term is largest. Nevertheless, both terms lead to a converging simulation and hence will be further tested going forward.

\subsection{Flow field with learned correction terms}
\begin{figure}[h]
\centering
\includegraphics[clip,trim={0.0\textwidth} {0.} {0.06\textwidth} {0.},width=\textwidth]{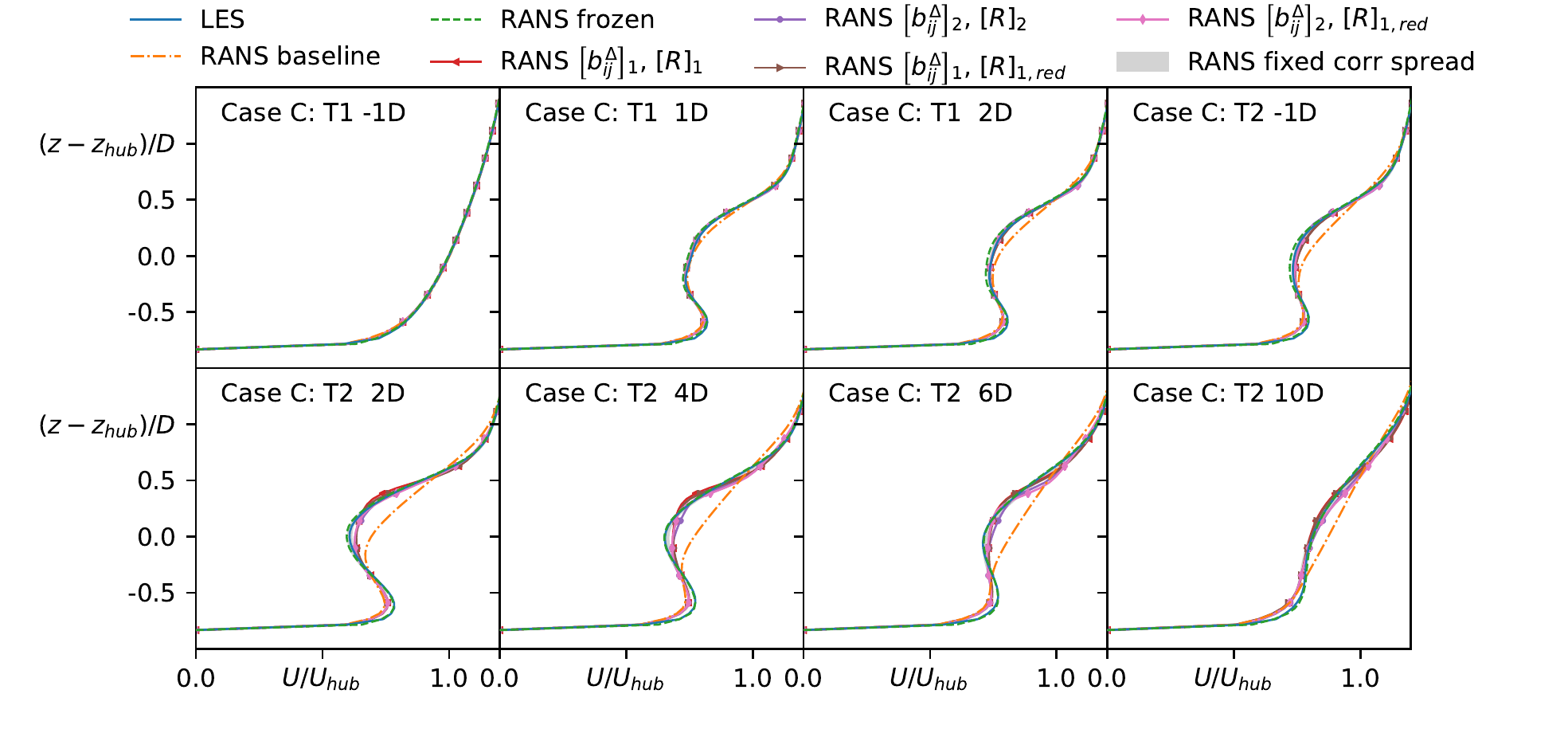}
\caption{\label{fig:caseCcoupledU} Comparison between LES, RANS baseline, frozen RANS and corrected RANS models via vertical slices of the velocity field up and downstream of the rotor plane for the three turbines of case C.}
\end{figure}

\begin{figure}[h]
\centering
\includegraphics[clip,trim={0.0\textwidth} {0.} {0.06\textwidth} {0.},width=\textwidth]{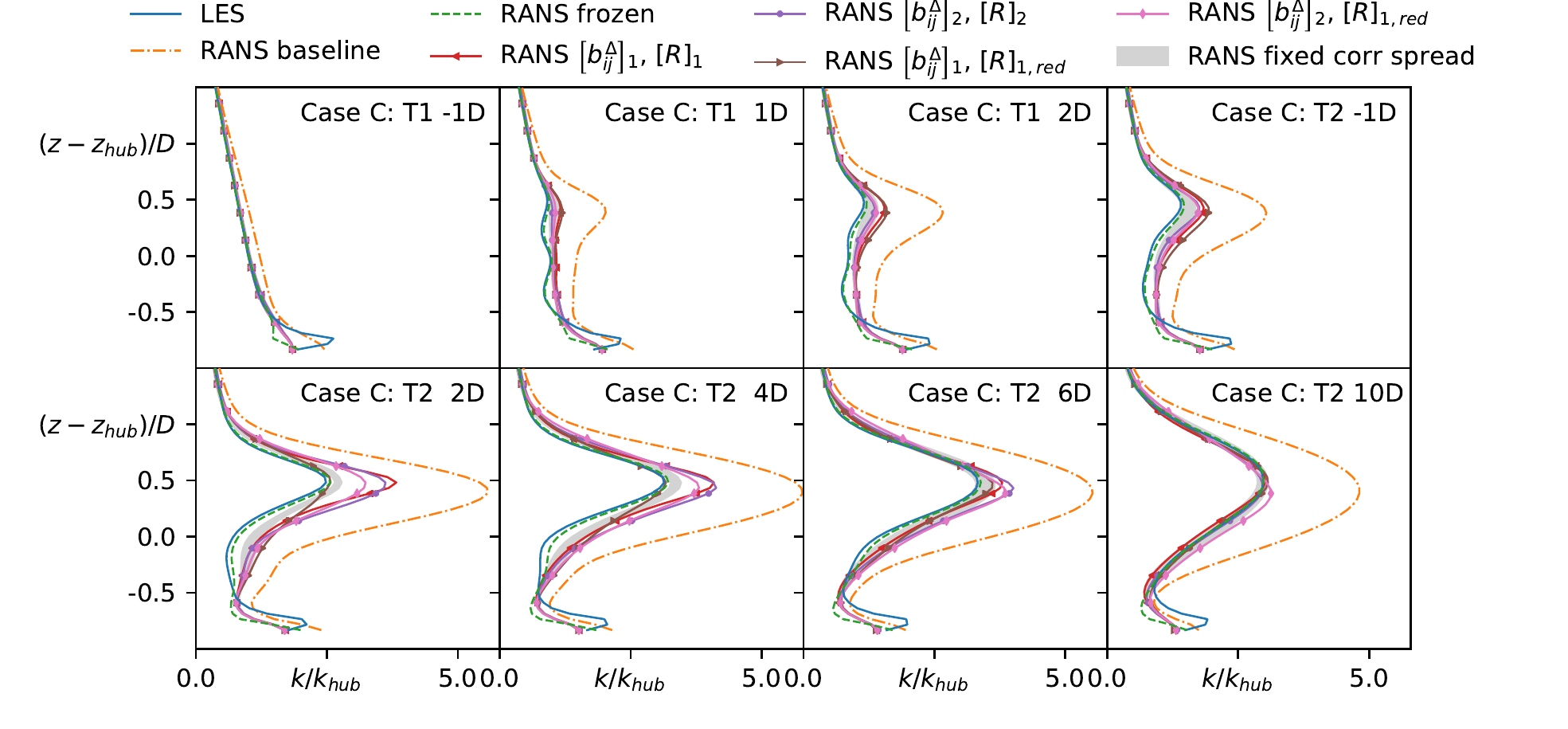}
\caption{\label{fig:caseCcoupledK} Comparison between LES, RANS baseline, frozen RANS and corrected RANS models via vertical slices of the turbulent kinetic energy field up and downstream of the rotor plane for the three turbines of case C.}
\end{figure}

Finally, now that model selection and assessment of model robustness have been carried out, both correction terms can be implemented simultaneously while coupled to the RANS flow field. The vertical profiles of velocity and $k$ are shown in Figures \ref{fig:caseCcoupledU} and \ref{fig:caseCcoupledK}.  For comparison, also the baseline model, the frozen case, and the flow field with fixed learned correction terms are shown. For the cases with the fixed correction terms, only the spread between all possible combinations of the correction terms is shown.

As is visible from the figures, the spread between the simulations with the fixed correction terms is smaller than the spread for the simulations where the correction terms are coupled to the RANS velocity field. This is quite logical given the results from the robustness analysis. Nevertheless, all the shown models yield a solid improvement over the baseline model in the wake region.  No results for the scalar correction term $\left[R\right]_{2,red}$ are shown because this term would lead to diverging simulations on both the test and training case, even when strong underrelaxation was used.

All the velocity profiles in figure \ref{fig:caseCcoupledU} from simulations with the coupled correction terms show significant improvement over the baseline model. In fact, the spread between the different models is minimal and the difference between the fixed and the coupled models is quite small. However, the discrepancy with respect to the reference profile increases further downstream akin to an error accumulation. Thus, it would be interesting to test the models on a case with more turbines to see how robust the enhanced models actually are.

In Figure \ref{fig:caseCcoupledK} the t.k.e.\ profiles are shown for the baseline and the improved models. Here, the spread between the coupled and the fixed models is larger, especially in the near wake of the second rotor. Comparison between the results for the various enhanced models indicates that the scalar correction term $R$ is what leads to the large spread between the models in the wake of the turbine. As compared to the velocity profiles, the discrepancy with respect to the reference does not increase downstream which is encouraging. There is also an unphysical underprediction of the t.k.e.\ close to the wall for the frozen case which is not present in the enhanced simulations: it seems the coupling with the flow field helps alleviate it. However, the t.k.e.\ close to the wall is still lower than the one for the reference time-averaged LES simulations, which show an unphysical overshoot there, so this discrepancy is actually a positive. The peak in the t.k.e.\ in the LES simulations is a well documented problem for LES simulations with wall functions for rough walls \cite{brasseur2010}. The plan is to address this in further publications.

According to the authors, overall the combination of the correction terms $\left[R\right]_{1,red}$ and $\left[b_{ij}^\Delta\right]_2$ yielded the best results and hence the full formulation for this correction terms is:
\begin{equation}
\label{eq:R}
\begin{split}
\left[R\right]_{1red} = 2 k \frac{\partial u_i}{\partial x_j} \ [ &
  \ \ \ \ 1.4771 \cdot 10^{-4} \cdot I_1^{0.5} \cdot q_\nu^{3.0} \cdot \textbf{T}^{(1)}_{ij}
  - 1.9183 \cdot q_{TI}^{0.5} \cdot q_{F}^{1.5} \cdot \textbf{T}^{(4)}_{ij} ] \\
+  \varepsilon \ [ &
\ \ \ \ 1.0970 \cdot 10^{1} \cdot q_{TI}^{0.5} \cdot q_F \cdot I_1^{0.5}
  + 6.1657 \cdot 10^{-5} \cdot q_{TI} \cdot I_1^{2.0} \cdot I_{34} \\
& + 8.3864 \cdot 10^{-3} \cdot q_{TI}^{1.5} \cdot I_{25}
 - 1.7888 \cdot 10^{2}   \cdot q_{TI}^{2.0} \cdot I_{25} \\
&  - 1.3956 \cdot 10^{1}   \cdot q_F \cdot q_\gamma^{0.5}
 + 2.5231 \cdot 10^{-7}  \cdot q_{TI}^{2.5} \cdot I_{25}^{2.0} \\
&  - 2.2330 \cdot q_F \cdot q_\gamma
 - 5.2367 \cdot 10^{-6} \cdot I_1^{2.0} \cdot q_\nu^{4.0} \\
&  - 5.5597 \cdot 10^{-2} \cdot q_\nu^{3.0}
 ]
\end{split}
\end{equation}
and
\begin{equation}
\label{eq:bDelta}
\begin{split}
\left[b_{ij}^\Delta\right]_2 = [
  &  \ \ \ \ 2.5095 \cdot 10^{-2} \cdot q_{TI}^{0.5} \cdot I_1^{0.5}
     + 1.090  \cdot 10^{-5} \cdot q_{TI} \cdot q_F^{0.5} \cdot I_1^{2.0} \\
  &  + 3.4089 \cdot 10^{-4} \cdot q_{TI}^{2.0} \cdot q_F^{0.5} \cdot I_1^{2.0}
     - 4.0175 \cdot 10^{-6} \cdot q_{TI}^{2.0} \cdot I_1^{2.0} \cdot q_{\nu} \\
  &  - 3.6356 \cdot 10^{-5} \cdot q_{TI}^{2.0} \cdot I_1^{2.5}
     + 9.6825 \cdot 10^{1}  \cdot q_{TI}^{3.0} \cdot q_{\nu}^{2.0} \\
  &  - 2.8904 \cdot 10^{3}  \cdot q_{TI}^{3.5}
     + 6.1482 \cdot 10^{-2} \cdot q_F^{0.5} \\
  &  - 9.4482 \cdot 10^{-5} \cdot q_F^{0.5} \cdot I_1 \cdot q_{\nu}^{2.0}
     - 2.1767 \cdot 10^{-3} \cdot q_{\nu}^{2.5} \\
  &  + 8.6126 \cdot 10^{-4} \cdot I_1^{0.5}
    ] \cdot  \textbf{T}^{(1)}_{ij}  \\
  + [
  &  - 9.4932 \cdot 10^{-2} \cdot q_{TI}^{0.5} \cdot q_F
     + 1.0716 \cdot 10^{-2} \cdot q_{TI}^{0.5} \cdot q_F^{1.5} \\
  &  + 6.3229 \cdot 10^{-4} \cdot q_{TI}^{0.5} \cdot q_{\nu}^{2.5}
     + 6.3233 \cdot 10^{-5} \cdot q_{TI}^{0.5} \cdot q_{\nu}^{3.0} \\
  &  + 3.7871 \cdot 10^{-4} \cdot q_{TI} \cdot I_{34}
     + 7.5746 \cdot 10^{-4} \cdot q_{TI}^{2.5} \cdot I_{18} \\
  &  - 1.7673 \cdot 10^{3}  \cdot q_{TI}^{4.5}
     + 4.8578 \cdot 10^{-3} \cdot q_F \\
  &  - 4.1741 \cdot 10^{-8} \cdot I_1^{0.5} \cdot I_2
     + 1.3261 \cdot 10^{-6} \cdot I_1
  ]  \cdot  \textbf{T}^{(2)}_{ij} \\
+ [
  &  - 1.3262 \cdot 10^{-3}
     - 2.7248 \cdot 10^{-6} \cdot I_1^{0.5} \cdot q_{\nu}^{4.0} \\
  &  + 6.5684 \cdot 10^{-7} \cdot I_1 \cdot q_{\nu}^{2.5}
  ] \cdot  \textbf{T}^{(3)}_{ij}  \\
  & - 3.5887 \cdot 10^{-5} \cdot q_{\nu}^{4.5} \cdot \textbf{T}^{(4)}_{ij}
\end{split}
\end{equation}

The anisotropy correction term $\left[b_{ij}^\Delta\right]_2$ consists of 25 terms of which 11 are multiplied by the first tensor of Pope's invariant basis, $\textbf{T}^{(1)} = \textbf{S}$, i.e.\ the correction tensor is linear.  Thus, this part of the correction tensor is implemented in the turbulence model in a semi-implicit form, and the remaining non-linear terms are incorporated in a fully explicit manner. The authors expect that this further increases the stability of the numerical implementation.

Some of the coefficients for the two correction models have a very small magnitude, so it may seem that they are not necessary. However, the influence of neglecting all coefficients was checked and the shown coefficients all result in a change in the relative mean or maximum error of at least 3 \% as compared to the full formulation shown above. Hence, all the shown terms are necessary even if the coefficient magnitude is small.

\subsection{Mesh convergence with learned correction terms}\label{sec:meshConvergence}
\begin{figure}[h]
\centering
\includegraphics[clip,trim={0.05\textwidth} {0.} {0.12\textwidth} {0.},width=\textwidth]{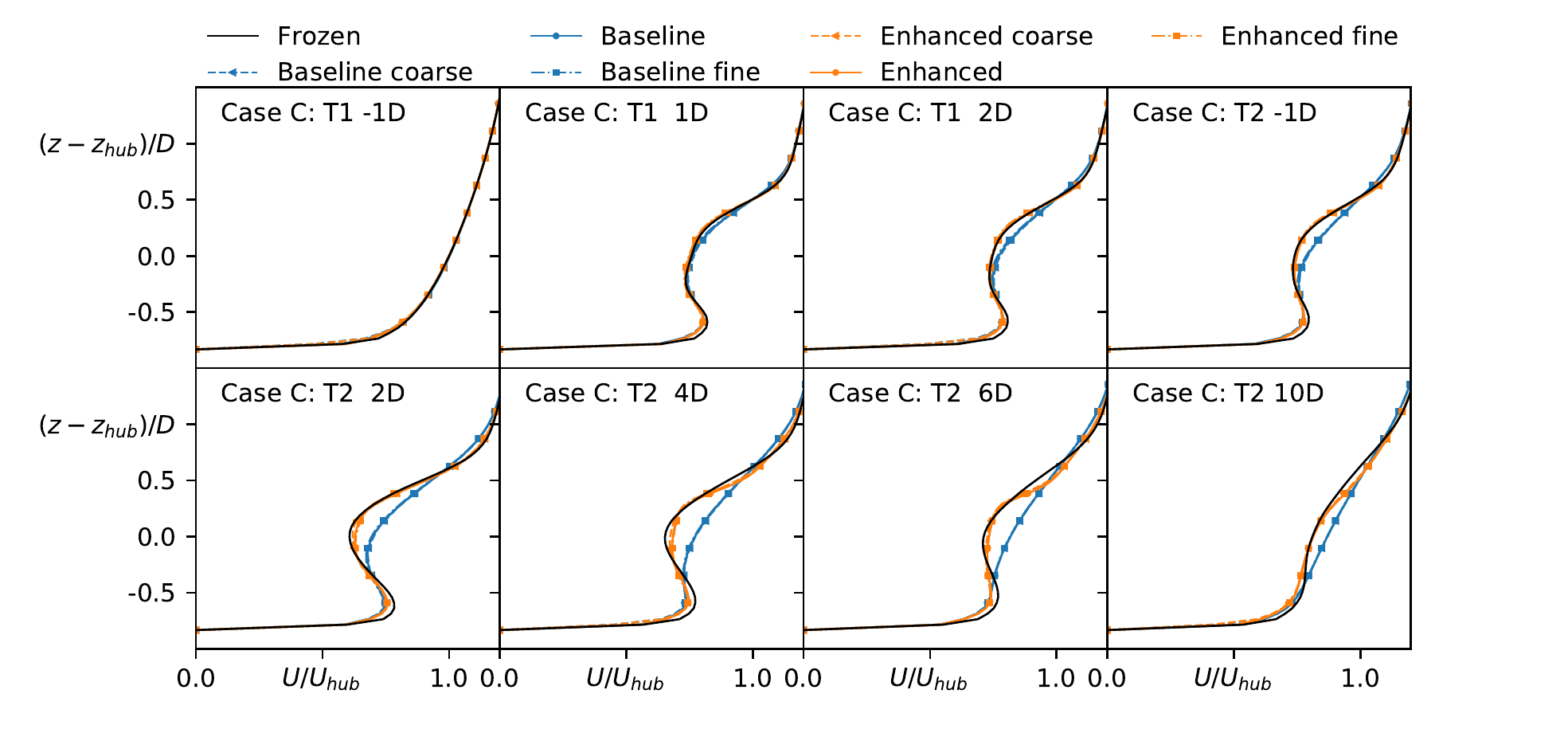}
\caption{\label{fig:meshCVu} Mesh convergence study for the baseline and the corrected model. Shown are the vertical slices of the velocity field up and downstream of the rotor plane for the three turbines of case C.}
\end{figure}

\begin{figure}[h]
\centering
\includegraphics[clip,trim={0.05\textwidth} {0.} {0.12\textwidth} {0.},width=\textwidth]{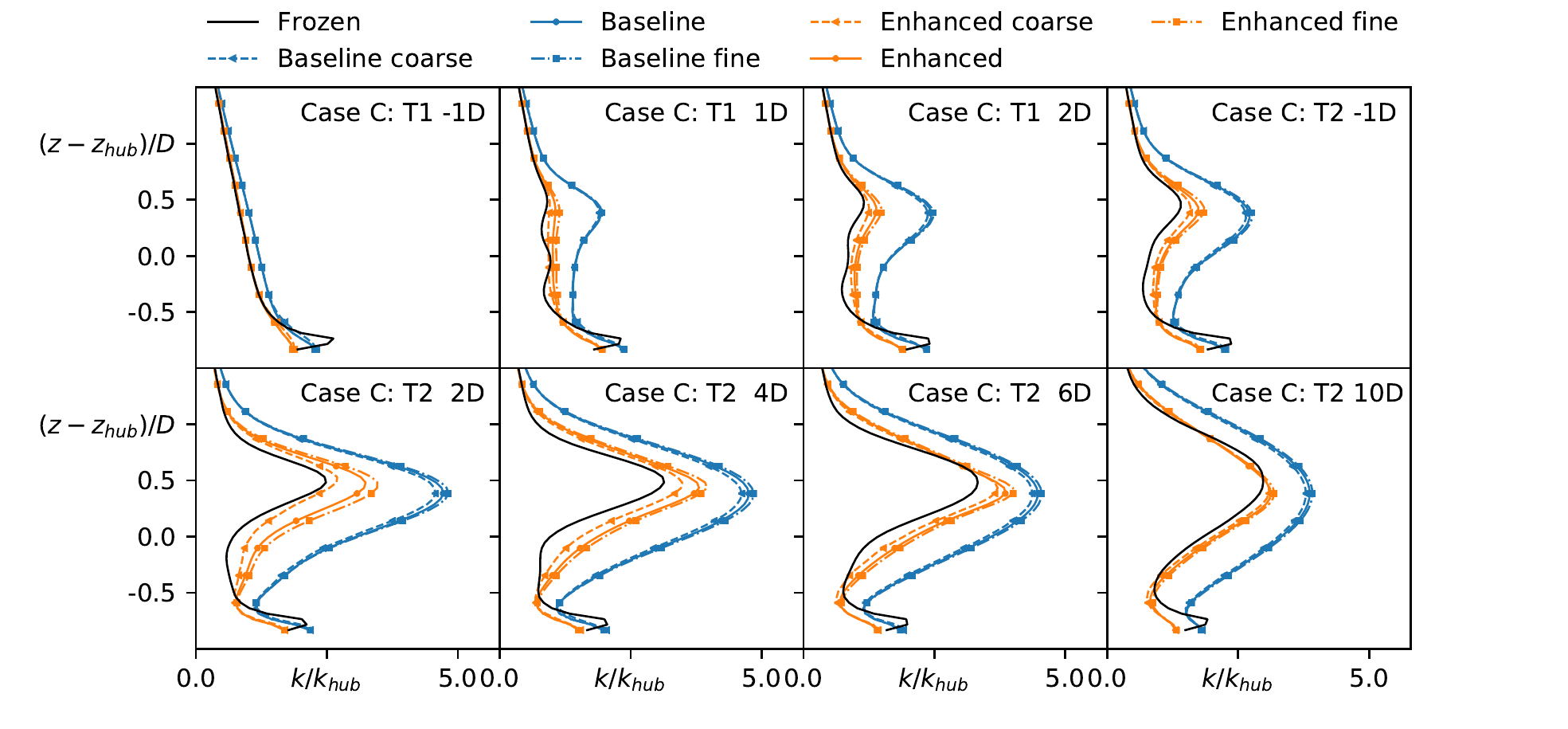}
\caption{\label{fig:meshCVk} Mesh convergence study for the baseline and the corrected model. Shown are the vertical slices of the turbulent kinetic energy field up and downstream of the rotor plane for the three turbines of case C.}
\end{figure}

As pointed out by Van Der Laan~\cite{laan2014}, nonlinear eddy viscosity models can be prone to numerical instability when a fine mesh is used.  To check whether the results of the developed model correction terms are actually grid-independent a mesh convergence study is carried out both for the baseline, as well as, the corrected model. The results for the vertical velocity and the turbulent kinetic energy profiles are shown in figures \ref{fig:meshCVu} and \ref{fig:meshCVk}.  The mesh properties are shown in Table \ref{tab:meshConvergence}.

\FloatBarrier

The velocity profiles in Figure \ref{fig:meshCVu} are insensitive to the mesh for both the baseline and the corrected models.  Hence, in terms of velocity the results are close to mesh independent at the presented refinement levels.
\begin{table}[h]
\centering
\begin{tabular}{@{}*{7}{l}}
\toprule \midrule
\textbf{Name} & \textbf{Density} $n_x \times n_y \times n_z$ \\
\midrule
Coarse & $240 \times 120 \times 48$\\
Reference (same as for LES) & $360 \times 120 \times 64$\\
Fine & $540 \times 240 \times 64$\\
\midrule \bottomrule
\end{tabular}
\caption{Mesh convergence parameters.}
\label{tab:meshConvergence}
\end{table}
There is more variation in the t.k.e., see Figure \ref{fig:meshCVk}, and the baseline model, shows less sensitivity than for the corrected model.  However, even for the corrected model the difference between the reference and the fine mesh is small indicating that the mesh is fine enough and that results are largely mesh independent.

Overall, these results are encouraging. The correction terms depend on the normalized rotor forcing which in turn depends on the actuator model, so there is an additional coupling loop in the prediction. Despite this there is little variation between the results.


\section{Conclusions}\label{sec:conclusions}
On a limited data set, the proposed frozen k-corrective frozen-RANS has demonstrate the potential for improving the predictions of the mean velocity and turbulent kinetic energy.  Based on time-averaged LES data-optimal correction terms to the turbulence transport equation of the RANS model were determined. The inclusion of these optimal correction terms in the RANS model leads to a near identical match between the RANS and the mean LES simulation. So this part of the methodology is working well. Then, the generalization of these terms through a sparse regression approach has yielded good results, but model selection is key. Some of the investigated models showed numerical instability once coupled with the RANS flow field. Hence, for future research the aim is to better understand the unstable coupling loops observed for some of the models which may be more likely to be triggered on fine meshes. Given this, either the structure of the correction terms can be changed or objective criteria for the model selection can be determined. Possibly also the inclusion of a classifier that helps distinguish between wake and not wake may help improve the accuracy of the correction terms or at least allow using simple formulations that are less prone to numerical instability. Further, the addition of an additional test case with more turbines would also help clarify what the limitations of the approach are.

Nevertheless, given that data-driven turbulence models are a relatively new development in the fluid dynamics community and so far they have only been applied to very fundamental cases, the obtained results are already a step in the right direction. Of course, to apply the methodology to more realistic conditions including atmospheric stratification and real scale turbines, a lot of work is still necessary.

\FloatBarrier

\appendix
\section{Physical features}\label{app:physicalFeatures}

\begin{table}[!htbp]
    \caption{Physics interpreted flow features. For each feature $q_i$ the physical description is denoted including the raw feature with its normalization. The features that are not Galilean invariant are marked with ${\dagger}$.}
	\begin{tabularx}{0.95\linewidth}{x b s s}
    \toprule \midrule
        ID & Description & Raw feature & Normalization\\ \midrule
  $q_Q$ & Ratio of excess rotation rate to strain rate (Q criterion)
    & $\frac{1}{2}(\left\|\boldsymbol{\Omega}\right\|^2 - \left\|\textbf{S}\right\|^2)$
    & $\left\|\textbf{S}\right\|^2$ \\
  $q_{TI}^{{\dagger}}$ & Turbulence intensity
    & $k$
    & $\frac{1}{2}U_iU_i$ \\
  $q_{ReD}$ & Wall distance based Reynolds number
    & $\frac{\sqrt{k}d}{50\nu}$
    &   - \\
  $q_{\partial p \partial s}^{{\dagger}}$ & Pressure gradient along streamline
    & $U_k\frac{\partial P}{\partial x_k}$
    & $\sqrt{\frac{\partial P}{\partial x_j}\frac{\partial P}{\partial x_j}U_i U_i}$ \\
  $q_T$   &   Ratio of mean turbulent to mean strain time scale
    & $\frac{k}{\varepsilon}$
    & $\frac{1}{\left\|\textbf{S}\right\|}$ \\
  $q_{\nu}$ & Viscosity ratio
      & $\nu_t$
      & $100\nu$ \\
  $q_{\perp}^{{\dagger}}$ & Nonorthogonality between velocity and its gradient
    & $|U_iU_j \frac{\partial U_i}{\partial x_j}|$
    & $\sqrt{U_lU_lU_i\frac{\partial U_i}{\partial x_j}U_k \frac{\partial U_k}{\partial x_j}}$ \\
  $q_{\mathcal{C}_k/\mathcal{P}_k}^{{\dagger}}$ & Ratio of convection to Boussinesq production of TKE
    & $U_i\frac{dk}{dx_i}$
    & $|\overline{u_j'u_k'}S_{jk}|$ \\
  $q_{\tau}$ & Ratio of total to normal Boussinesq Reynolds stresses
    & $||\overline{u_i'u_j'}_{BS}||$
    &  $k$ \\
  $q_{\gamma}$ & Shear parameter
    & $\left\|\frac{\partial U_i}{\partial x_j}\right\|$
    & $\frac{\varepsilon}{k}$ \\
  $q_{F}^{{\dagger}}$ & Actuator forcing
  	& $\left\|F_{cell}\right\|$
    & $\frac{1}{2} \rho_0 A_{cell} \left\| U \right\|^2$ \\
	\midrule \bottomrule
	\end{tabularx}
    \label{tab:physicalFeaturesAll}
\end{table}

\FloatBarrier

\section{Integrity basis and invariants}\label{app:integrityBasis}
\begin{table}[!htbp]
    \centering
    \caption{Invariant bases, number of symmetric and antisymmetric tensors for each invariant are indicated by $n_s$ and $n_A$, respectively. The invariant bases are the trace of the tensors listed. The asterisk on a invariant bases indicates that also the cyclic permutation of the antisymmetric tensors are included.}
    \def\arraystretch{1.25}
    \begin{tabular}{ccc} \toprule[1pt]\midrule[0.3pt]
        $(n_S,n_A)$ & Feature index & Invariant bases \\ \midrule
        $(1,0)$   &   1-2 &   $\textbf{S}^2$, $\textbf{S}^3$    \\
        $(0,1)$   &   3-5 &   $\boldsymbol{\Omega}^2$, $\textbf{A}_p^2$, $\textbf{A}_k^2$      \\
        $(1,1)$   &   6-14 &   $\boldsymbol{\Omega}^2 \textbf{S}$, $\boldsymbol{\Omega}^2 \textbf{S}^2$, $\boldsymbol{\Omega}^2 \textbf{S} \boldsymbol{\Omega} \textbf{S}^2$   \\
           &    &   $\textbf{A}_p^2\textbf{S}$, $\textbf{A}_p^2\textbf{S}^2$, $\textbf{A}_p^2 \textbf{S} \textbf{A}_p \textbf{S}^2$    \\
           &   & $\textbf{A}^2_k\textbf{S}$, $\textbf{A}^2_k\textbf{S}^2$ , $\textbf{A}^2_k\textbf{S}\textbf{A}_k\textbf{S}^2$ \\
        $(0,2)$   &   15-17 &   $\boldsymbol{\Omega}\textbf{A}_p$, $\textbf{A}_p\textbf{A}_k$, $\boldsymbol{\Omega}\textbf{A}_k$   \\
        $(1,2)$   &   18-41 &   $\boldsymbol{\Omega}\textbf{A}_p\textbf{S}$, $\boldsymbol{\Omega}\textbf{A}_p\textbf{S}^2$, $\boldsymbol{\Omega}^2 \textbf{A}_p\textbf{S}^*$, $\boldsymbol{\Omega}^2\textbf{A}_p\textbf{S}^{2*}$, $\boldsymbol{\Omega}^2\textbf{S}\textbf{A}_p\textbf{S}^{2*}$   \\
           &    &   $\boldsymbol{\Omega}\textbf{A}_k\textbf{S}$, $\boldsymbol{\Omega}\textbf{A}_k\textbf{S}^2$, $\boldsymbol{\Omega}^2 \textbf{A}_k\textbf{S}^*$, $\boldsymbol{\Omega}^2\textbf{A}_k\textbf{S}^{2*}$, $\boldsymbol{\Omega}^2\textbf{S}\textbf{A}_k\textbf{S}^{2*}$    \\
           &   &    $\textbf{A}_p\textbf{A}_k\textbf{S}$, $\textbf{A}_p\textbf{A}_k\textbf{S}^2$, $\textbf{A}^2_p\textbf{A}_k\textbf{S}^*$, $\textbf{A}^2_p\textbf{A}_k\textbf{S}^{2*}$  \\
        $(0,3)$   &   42 &   $\boldsymbol{\Omega}\textbf{A}_p \textbf{A}_k$  \\
        $(1,3)$ & 43-47 & $\boldsymbol{\Omega}\textbf{A}_p\textbf{A}_k\textbf{S}$, $\boldsymbol{\Omega}\textbf{A}_k\textbf{A}_p\textbf{S}$, $\boldsymbol{\Omega}\textbf{A}_p\textbf{A}_k\textbf{S}^2$, $\boldsymbol{\Omega}\textbf{A}_k\textbf{A}_p\textbf{S}^2$, $\boldsymbol{\Omega}\textbf{A}_p\textbf{S}\textbf{A}_k\textbf{S}^2$ \\ \midrule[0.3pt]\bottomrule[1pt]
    \end{tabular}
    \label{tab:invariantsAll}
\end{table}

\FloatBarrier

\section{Horizontal slices}\label{app:horizontalSlices}

\begin{figure}[h]
\centering
\includegraphics[clip,trim={0.05\textwidth} {0.} {0.12\textwidth} {0.},width=\textwidth]{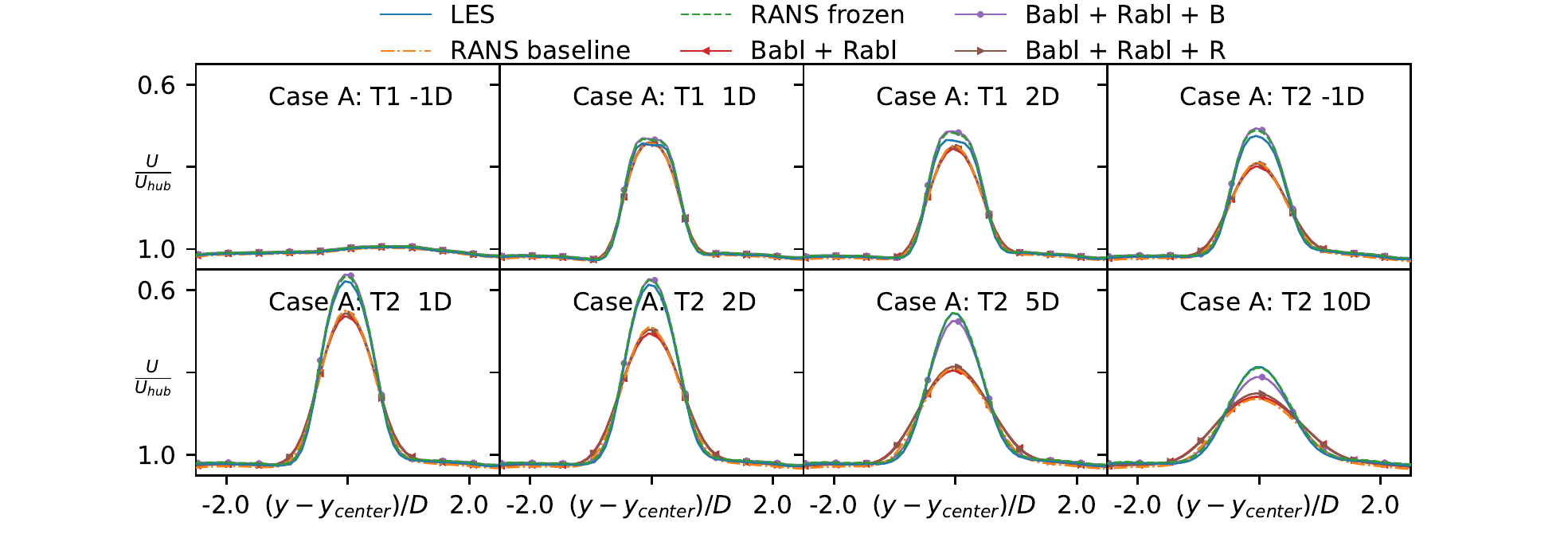}
\caption{\label{fig:modelFrozenUhor} Comparison between LES, RANS baseline and frozen RANS with selective inclusion of the different components of the correction terms via horizontal slices of the velocity field up and downstream of the rotor plane of the two turbines of case A.}
\end{figure}

\begin{figure}
\includegraphics[clip,trim={0.05\textwidth} {0.} {0.12\textwidth} {0.},width=\textwidth]{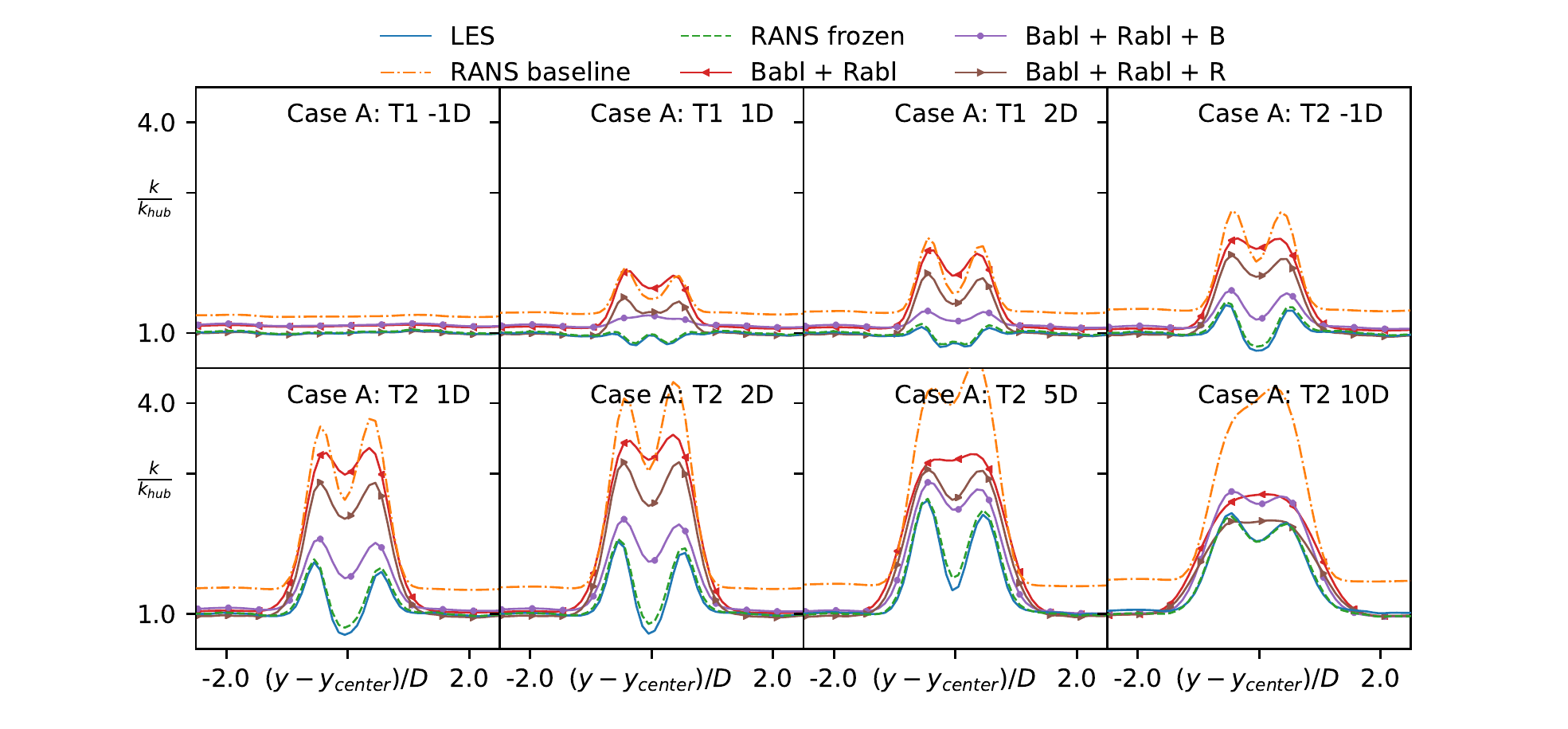}
\caption{\label{fig:modelFrozenKhor} Comparison between LES, RANS baseline and frozen RANS with selective inclusion of the different components of the correction terms via horizontal slices of the turbulent kinetic energy field up and downstream of the rotor plane of the two turbines of case A.}
\end{figure}

\begin{figure}[h]
\centering
\includegraphics[clip,trim={0.05\textwidth} {0.} {0.08\textwidth} {0.},width=\textwidth]{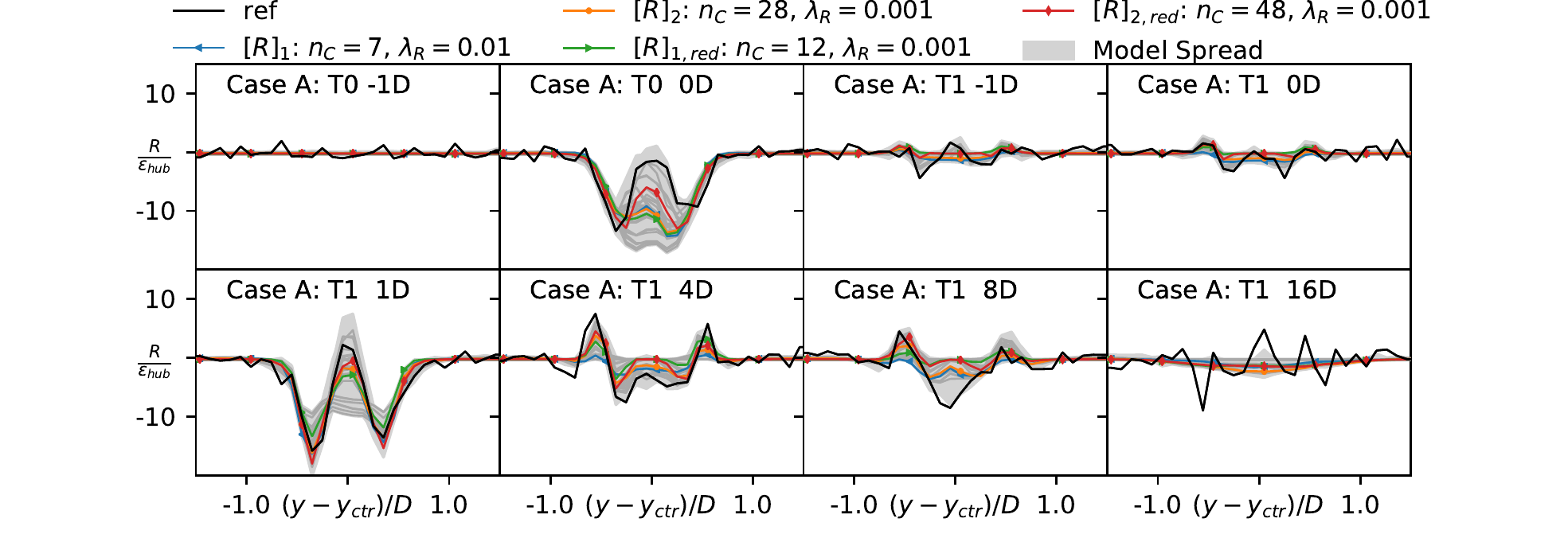}
\caption{\label{fig:modelSpreadRhor} Spread of learned model correction $R$ for case A through horizontal slices at the rotor plane at different streamwise stations as labeled in the subplots. The model spread is for all models that are Pareto optimal as defined previously. The models selected during the cliqueing post-processing step are shown explicitly either in color or in dark gray. The models selected for further investigation are highlighted in color. Finally, the optimal correction term is shown in black.}
\end{figure}

\begin{figure}[h]
\centering
\includegraphics[clip,trim={0.05\textwidth} {0.} {0.06\textwidth} {0.},width=\textwidth]{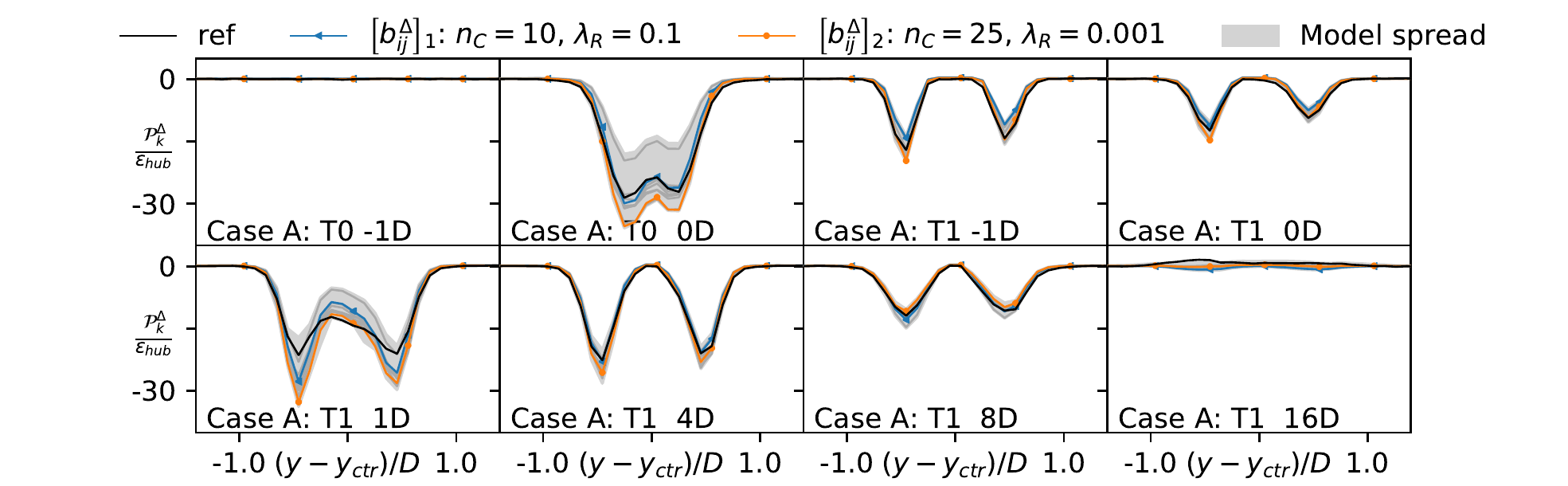}
\caption{\label{fig:modelSpreadPkDeltaHor} Spread of learned model correction $\mathcal{P}_{k}^\Delta$ for case A through horizontal slices at the rotor plane at different streamwise stations as labeled in the subplots. The model spread is for all models that are Pareto optimal as defined previously. The models selected during the cliqueing post-processing step are shown explicitly either in color or in dark gray. The models selected for further investigation are highlighted in color. Finally, the optimal correction term is shown in black. }
\end{figure}

\begin{figure}[h]
\centering
\includegraphics[clip,trim={0.05\textwidth} {0.} {0.0\textwidth} {0.},width=\textwidth]{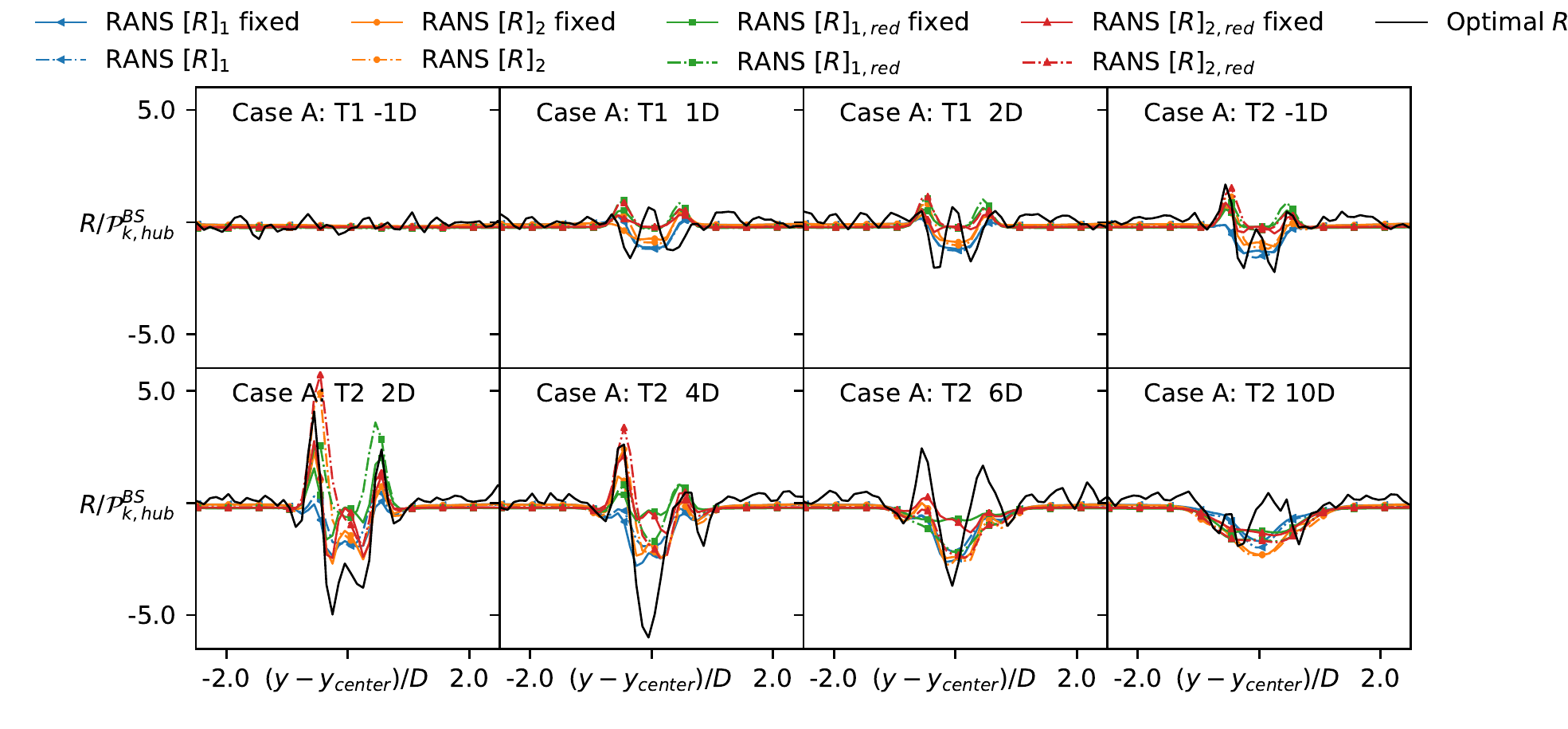}
\caption{\label{fig:robustnessRhor} Spread of fixed and coupled corrections $R$ for the training case A via horizontal slices at the rotor plane up and downstream of the two turbines.}
\end{figure}

\begin{figure}[h]
\centering
\includegraphics[clip,trim={0.05\textwidth} {0.} {0.0\textwidth} {0.},width=\textwidth]{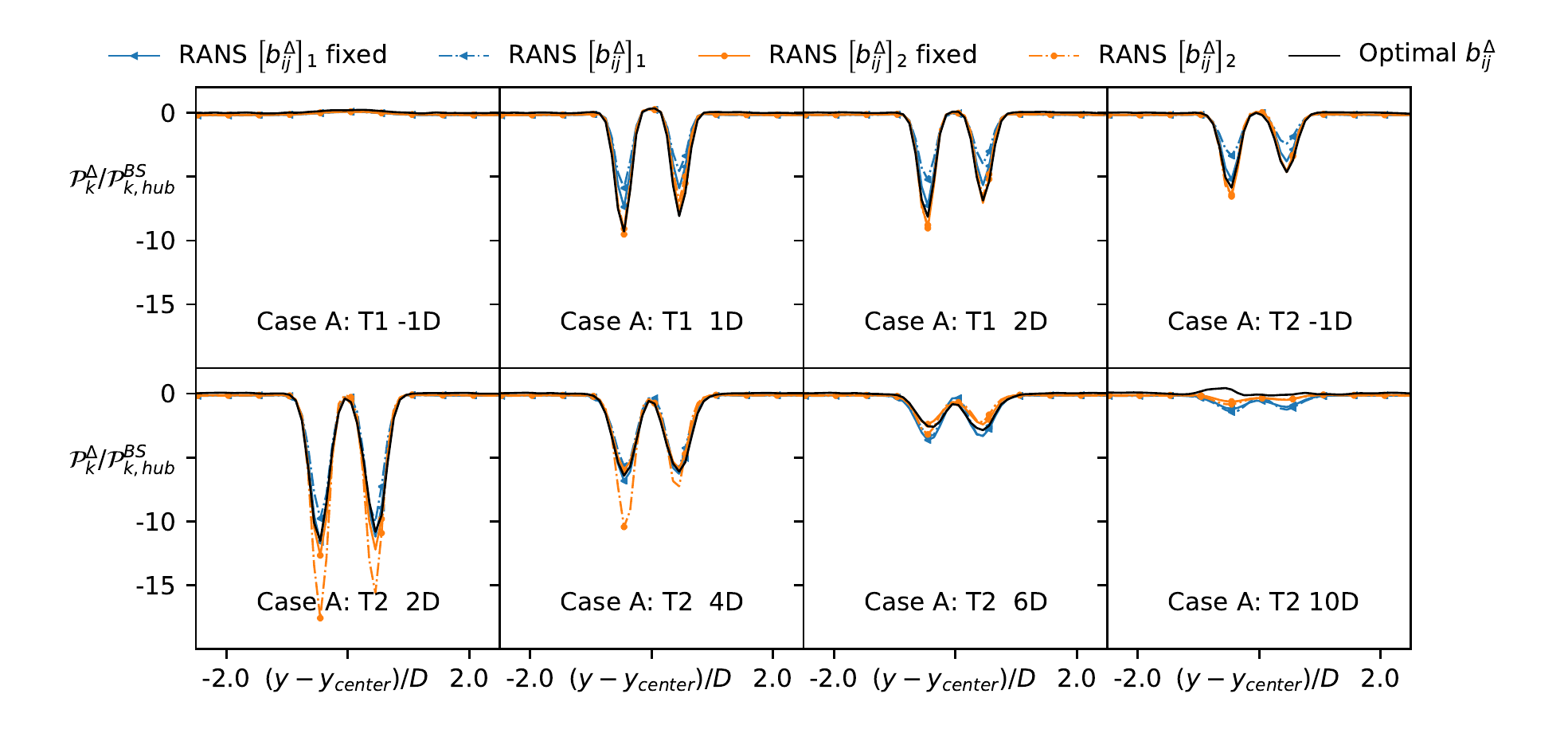}
\caption{\label{fig:robustnessPkDeltaHor} Spread of fixed and coupled corrections $b_{ij}^\Delta$ in terms of  $\mathcal{P}_{k}^\Delta$ for the training case A via horizontal slices at the rotor plane up and downstream of the two turbines.}
\end{figure}

\begin{figure}[h]
\centering
\includegraphics[clip,trim={0.05\textwidth} {0.} {0.0\textwidth} {0.},width=\textwidth]{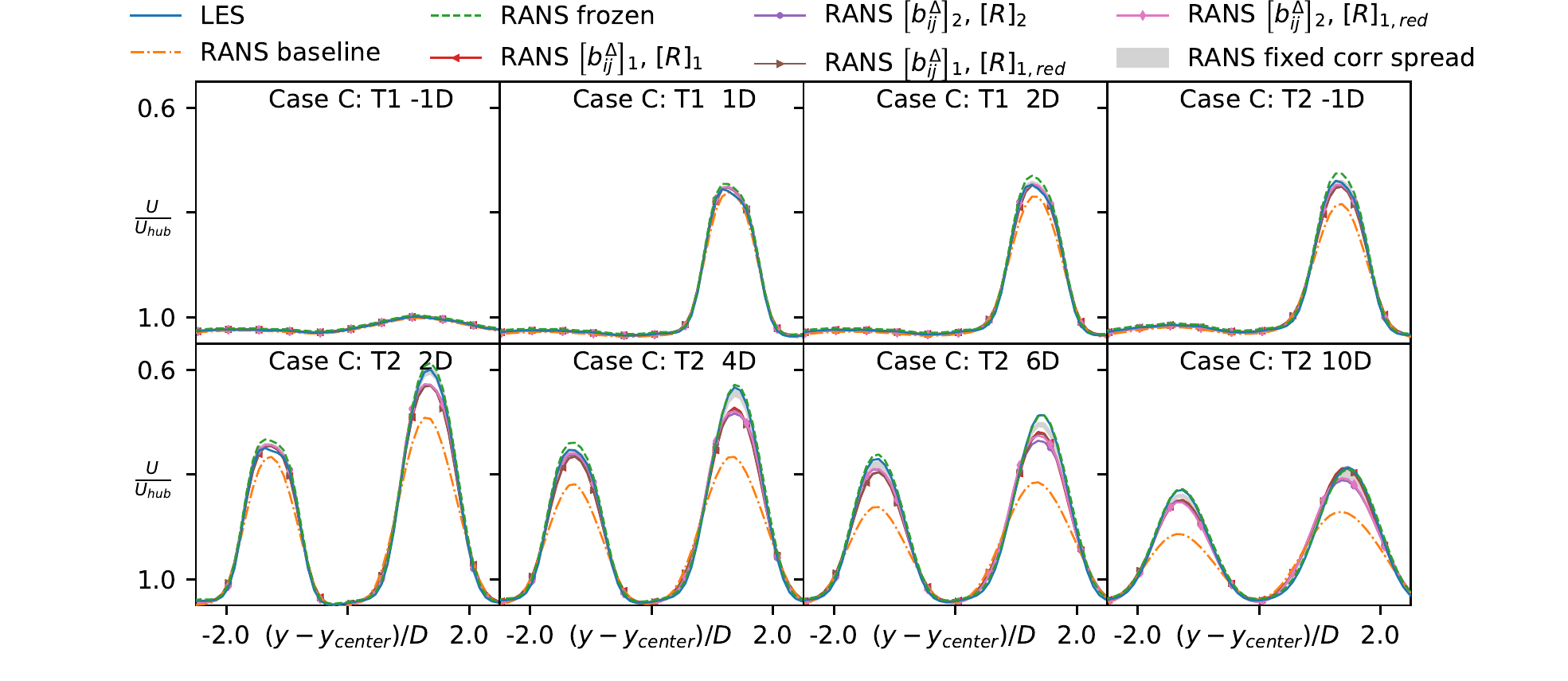}
\caption{\label{fig:caseCcoupledUhor} Comparison between LES, RANS baseline, frozen RANS and corrected RANS models via horizontal slices of the velocity field up and downstream of the rotor plane for the three turbines of case C.}
\end{figure}

\begin{figure}[h]
\centering
\includegraphics[clip,trim={0.05\textwidth} {0.} {0.08\textwidth} {0.},width=\textwidth]{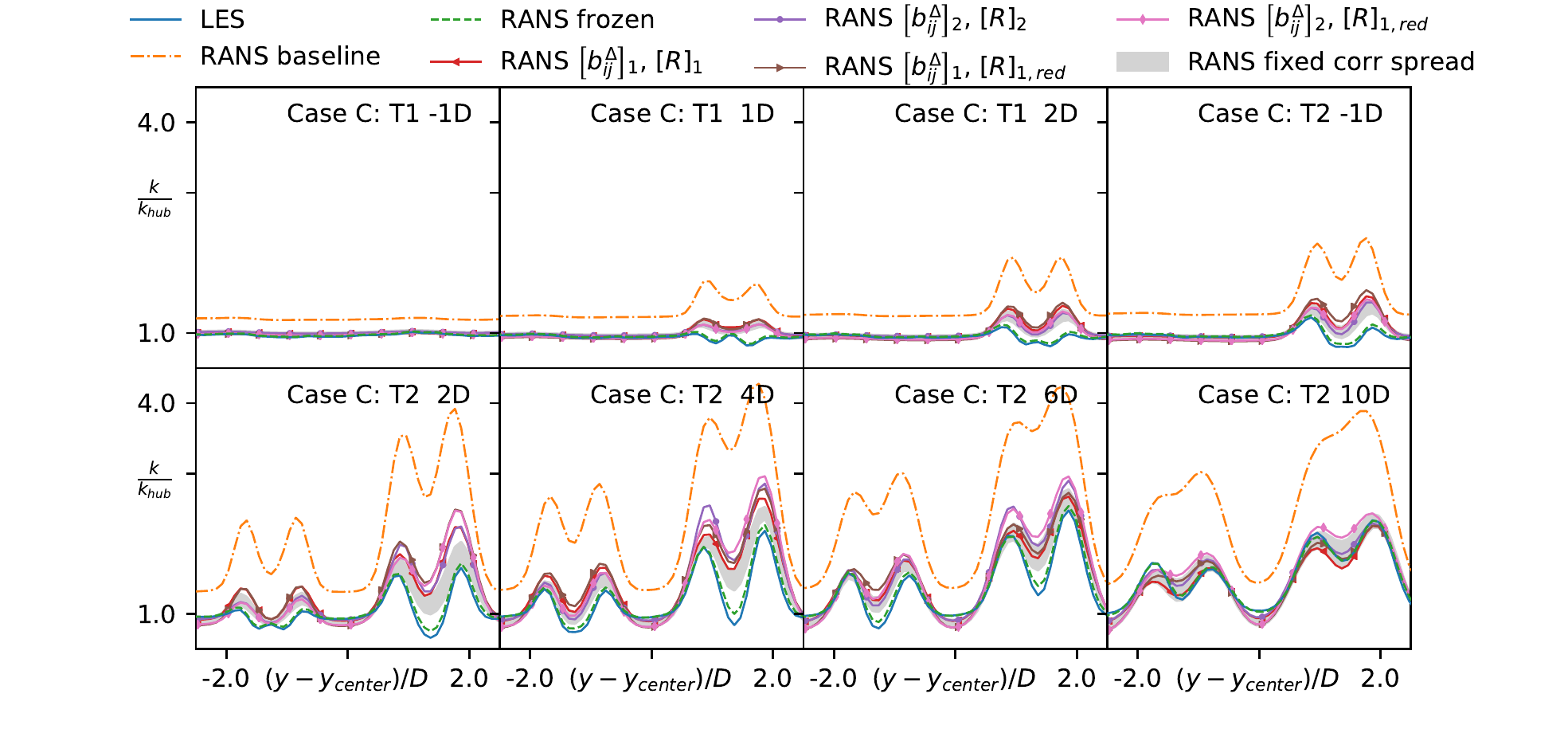}
\caption{\label{fig:caseCcoupledKhor} Comparison between LES, RANS baseline, frozen RANS and corrected RANS models via horizontal slices of the turbulent kinetic energy field up and downstream of the rotor plane for the three turbines of case C.}
\end{figure}

\begin{figure}[h]
\centering
\includegraphics[clip,trim={0.05\textwidth} {0.} {0.06\textwidth} {0.},width=\textwidth]{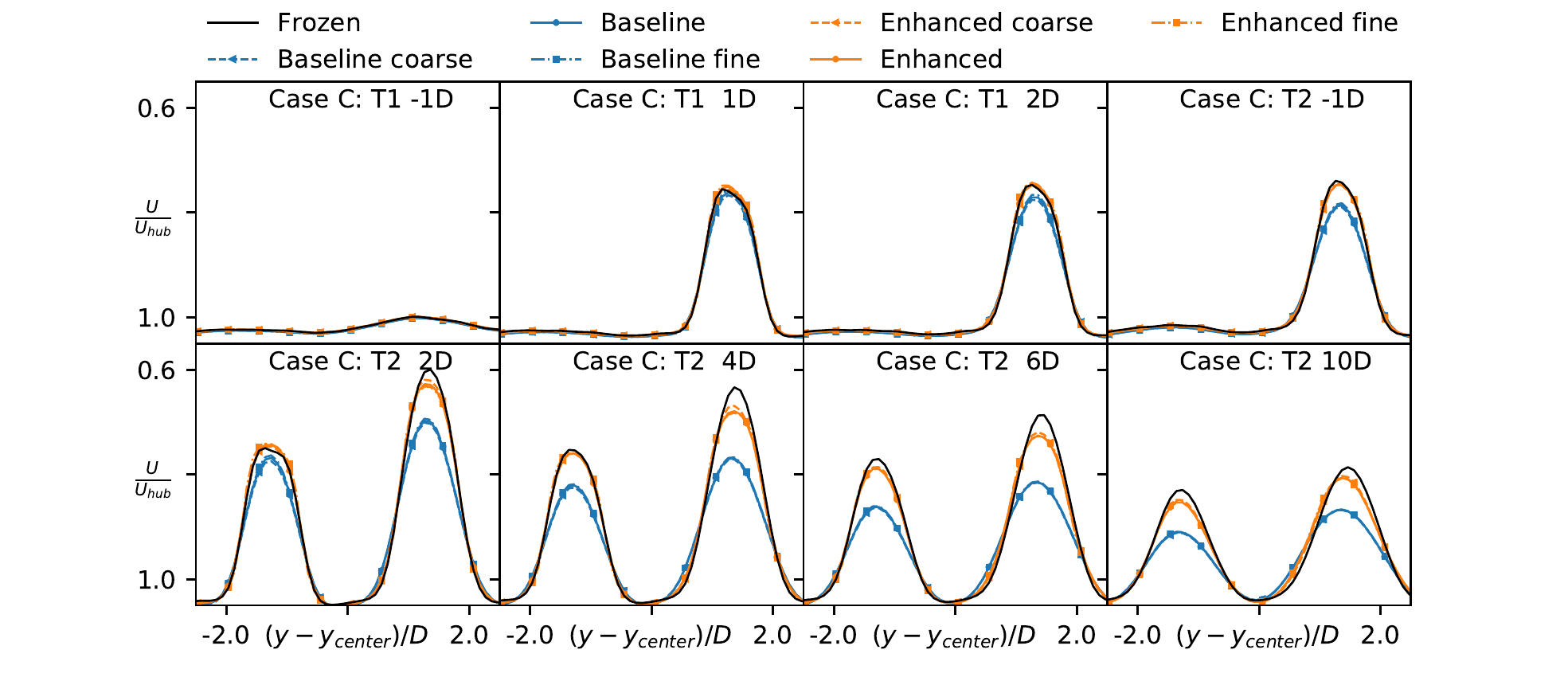}
\caption{\label{fig:meshCVuHor} Mesh convergence study for the baseline and the corrected model. Shown are the horizontal slices of the velocity field up and downstream of the rotor plane for the three turbines of case C.}
\end{figure}

\begin{figure}[h]
\centering
\includegraphics[clip,trim={0.05\textwidth} {0.} {0.12\textwidth} {0.},width=\textwidth]{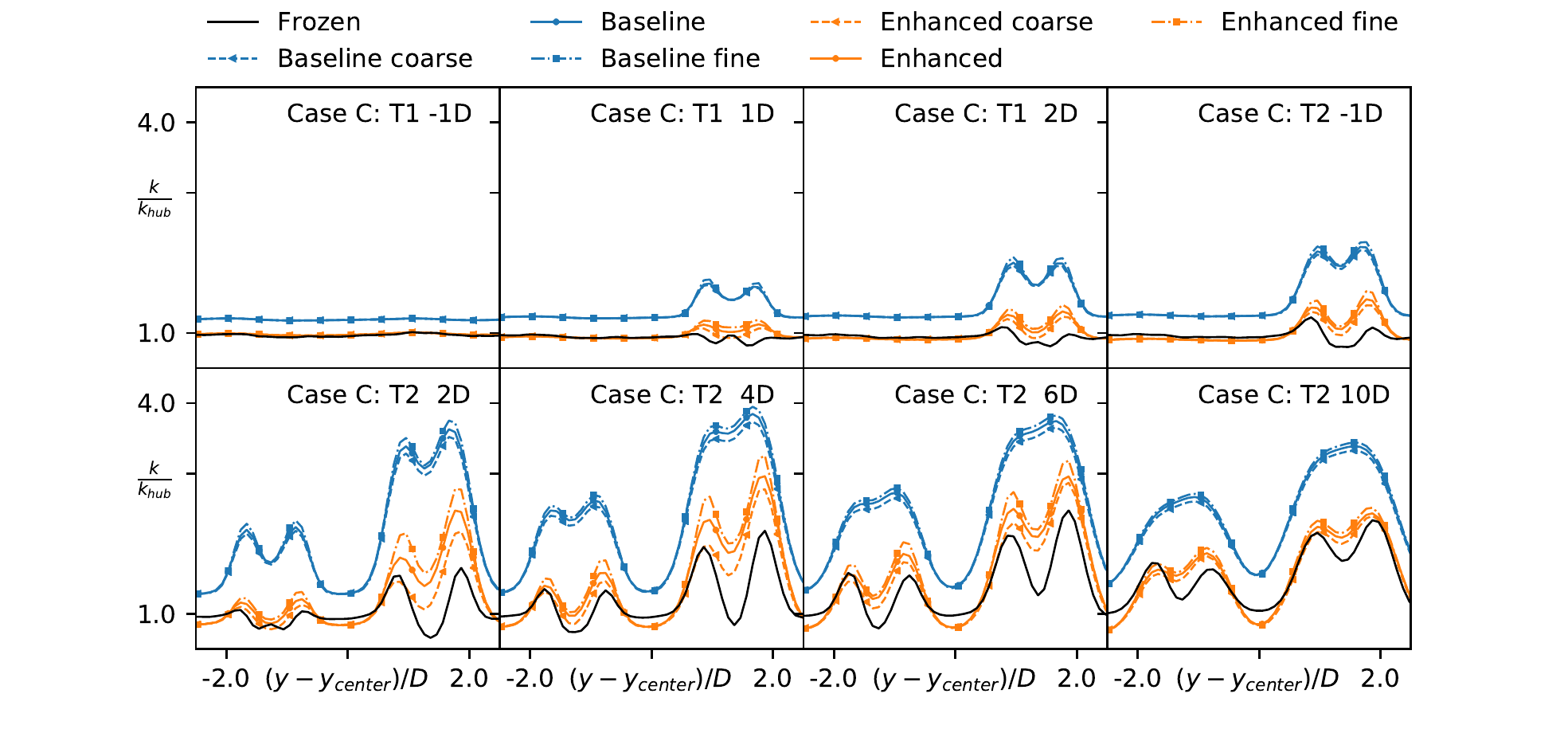}
\caption{\label{fig:meshCVkHor} Mesh convergence study for the baseline and the corrected model. Shown are the horizontal slices of the turbulent kinetic energy field up and downstream of the rotor plane for the three turbines of case C.}
\end{figure}

\FloatBarrier

\bibliography{bibl}

\end{document}